%% file: main.tex
\documentclass[sigconf]{acmart}

\usepackage{multirow}
\usepackage[inline]{enumitem}
\usepackage[most]{tcolorbox}
\usepackage[normalem]{ulem}
\usepackage{xcolor}
\usepackage{lipsum}
\usepackage[skip=2pt,labelfont=bf]{caption}
\newcommand{\find}[1]{
\begin{tcolorbox}[leftrule=1mm,toprule=0mm,bottomrule=0mm,left=1pt,right=2pt,top=2pt,bottom=2pt]%[tile,size=fbox,boxsep=2mm,boxrule=0pt,top=0pt,bottom=0pt,borderline={0.5mm}{0pt}{black!70!white},colback=black!5!white]
#1
\end{tcolorbox}
}

\definecolor{findgray}{gray}{0.95}

\newenvironment{findingbox}{%
  \def\FrameCommand##1{%
    \setlength{\fboxsep}{6pt}%    Distance between text and border (padding)
    \setlength{\fboxrule}{1.5pt}% Thickness of the vertical lines
    \textcolor{black}{\vrule width \fboxrule}%  Left Vertical Line
    \colorbox{findgray}{##1}%                   Background Box
    \textcolor{black}{\vrule width \fboxrule}%  Right Vertical Line
  }%
  \MakeFramed {\advance\hsize-\width \FrameRestore}}%
 {\endMakeFramed}

\newcommand{\paragraphbold}[1]{\noindent\textbf{#1}}
\newcommand{\subsubsectionbold}[1]{\subsubsection{\textbf{#1}}}
\newcommand{\paragraphtight}[1]{\textit{#1}}

\newcommand{\hypobox}[1]{%
\begin{tcolorbox}[
  enhanced,
  colback=white,
  colframe=black,
  boxrule=0.8pt,
  left=6pt,
  right=6pt,
  top=4pt,
  bottom=3pt,
  boxsep=2pt,
  width=\columnwidth,
]
\bfseries #1
\end{tcolorbox}%
}

\AtBeginDocument{%
  }

\renewcommand\footnotetextcopyrightpermission[1]{}
\acmConference{}{}{}
\acmBooktitle{}
\acmDOI{}
\acmISBN{}

\newcommand{\agentlessname}{Agentless}
\newcommand{\kgcompassname}{KGCompass}
\newcommand{\experepairname}{ExpeRepair}
\newcommand{\agentless}{\textsc{\agentlessname}}
\newcommand{\kgcompass}{\textsc{\kgcompassname}}
\newcommand{\experepair}{\textsc{\experepairname}}
\newcommand{\swebench}{SWE-bench}
\newcommand{\swebenchlite}{\swebench\ Lite}
\newcommand{\bestofktext}{Best-of-K}
\newcommand{\bestofk}{Best-of-$K$}

\begin{document}

\title{Beyond Localization: Recoverable Headroom and Residual Frontier in Repository-Level RAG-APR}

\input{authors}

\renewcommand{\shortauthors}{Zhao et al.}

\input{abstract}

\maketitle

\input{introductionv2}
\input{backgroundv2}
\input{relatedworkv2}
\input{experimental_setupv2}

% \ \textcolor{blue}{New experimental data are being integrated and there will be significant changes}
\input{resultsv2}
\input{analysisv2}
\input{conclusionv2}
% writing
% \include{discussion}

%%
%% The next two lines define the bibliography style to be used, and
%% the bibliography file.
\bibliographystyle{ACM-Reference-Format}
\bibliography{main}

\end{document}

%% file: authors.tex
\author{Pengtao Zhao}
\affiliation{%
  \institution{School of Computing and Information Systems, University of Melbourne}
  \country{Australia}
}
\email{pengtaozhao@unimelb.edu.au}

\author{Boyang Yang}
\affiliation{%
  \institution{School of Artificial Intelligence (School of Software), Yanshan University}
  \country{China}
}
\email{yby@ieee.org}

\author{Bach Le}
\affiliation{%
  \institution{School of Computing and Information Systems, University of Melbourne}
  \country{Australia}
}
\email{bach.le@unimelb.edu.au}

\author{Feng Liu}
\affiliation{%
  \institution{School of Computing and Information Systems, University of Melbourne}
  \country{Australia}
}
\email{feng.liu1@unimelb.edu.au}

\author{Haoye Tian}
\authornote{Corresponding author.}
\affiliation{%
  \institution{Department of Computer Science, Aalto University}
  \country{Finland}
}
\email{haoye.tian@aalto.fi}

%% file: abstract.tex
\begin{abstract}
    Repository-level automated program repair (APR) increasingly treats stronger localization as the main path to better repair.
    We ask a more targeted question: once localization is strengthened, which post-localization levers still provide recoverable gains, which are bounded within our protocol, and what residual frontier remains?
    We study this question on \swebenchlite{} with three representative repository-level RAG-APR paradigms, \agentless{}, \kgcompass{}, and \experepair{}.
    Our protocol combines Oracle Localization, within-pool \bestofk{}, fixed-interface added context probes with per-condition same-token filler controls and same-repository hard negatives, and a common-wrapper oracle check.
    Oracle Localization improves all three systems, but Oracle success still stays below 50\%.
    Extra candidate diversity still helps inside the sampled 10-patch pools, but that headroom saturates quickly.
    Under the two fixed interfaces, most informative added context conditions still outperform their own matched controls.
    The common-wrapper check shows different system responses: under a common wrapper, gains remain large for \kgcompass{} and \experepair{}, while \agentless{} changes more with builder choice.
    Prompt-level fusion still leaves a large residual frontier: the best fixed probe adds only 6 solved instances beyond the native three-system \textsc{Solved@}10 union.
    Overall, stronger localization, bounded search, evidence quality, and interface design all shape repository-level repair outcomes.
\end{abstract}

%% file: introductionv2.tex
\section{Introduction}
\label{sec:intro}

Repository-level automated program repair (APR) on benchmarks such as \swebench{} \cite{SWE-bench,swebench_lite_website} requires large language models (LLMs) to work over codebases that are much larger than one prompt.
Recent systems therefore rely on retrieval-augmented generation (RAG) and vector indexes such as Faiss \cite{lewis2020_rag,ref37_faiss_2024} to build bounded, task-relevant repair contexts.
We use \emph{retrieval-augmented generation based automated program repair} (RAG-APR) to refer to repository-level APR systems that retrieve bounded task-relevant evidence before patch generation.
Across current repository-level repair systems, this bounded-context RAG formulation has become a mainstream practical route, including phase-structured pipelines, structure-aware retrieval, memory-augmented repair, and software-agent style systems \cite{SWE-Agent,zhang2024_autocoderover,ref20_openhands_2024,Agentless,KGCompass,ExpeRepair,ref23_sgagent_2026,ref25_reporepair_2026}.

This success has also encouraged a common optimization story: localization first \cite{Agentless,SWE-Agent}.
If retrieval can find the right files and spans, repair should improve substantially, and recent localization-oriented analyses support that intuition \cite{Liu2019FLBias,wu2023largelanguagemodelsfault,Kang2024LLMExplainableFaultLocalization,10.1007/s10515-025-00549-x,RGFL2026}.
However, current RAG-APR papers still mainly compare end-to-end solve rates or new retrieval designs \cite{zhang2024_autocoderover}.
Those evaluations do not cleanly separate gains from localization, context construction, patch generation, and validation \cite{ref08_dissecting_leaderboards_2025,ref09_whats_in_benchmark_2026,ContextBench2026}.
As a result, they leave open a more important question: after localization gets stronger, what remains the main source of recoverable gain?
This is not just an evaluation-detail issue: without isolating the post-localization bottlenecks, apparent gains can be misattributed to retrieval itself, even when the limiting factor lies in how evidence is packaged, consumed, or translated into valid patches \cite{ref08_dissecting_leaderboards_2025,ContextBench2026,liu2023_lostinthemiddle}.
That distinction matters for both system design and scientific interpretation, because it determines whether further progress should come from better localization, better interface construction, or stronger patch-generation and validation strategies \cite{ref09_whats_in_benchmark_2026,RepoFixEval2026}.

There are already useful empirical studies and technical analyses around repository-level repair.
Benchmark audits examine leaderboard composition, model memory, solved-issue correctness, and evaluation reliability, while context studies show that long context is not automatically useful and that retrieved evidence can still be used poorly \cite{ref07_swebench_memory_2025,ref08_dissecting_leaderboards_2025,ref09_whats_in_benchmark_2026,ref10_rigorous_eval_agents_2025,ref11_solved_issues_correctness_2025,ref12_saving_swebench_2025,ContextBench2026,liu2023_lostinthemiddle}.
Repository-level task studies also argue for finer-grained decomposition of issue resolution \cite{RLCE,RepoFixEval2026}.
These papers make the field better understood, but they still stop short of one controlled post-localization study that measures three things together: how much within-system search headroom remains, whether added cross-paradigm evidence still helps under controlled interfaces, and how much frontier remains jointly unsolved after those gains are exhausted \cite{RGFL2026,ContextBench2026,RepoFixEval2026}.
Without that joint view, it is still hard to tell whether current repository-level repair systems are primarily limited by finding the right evidence, by using already-retrieved evidence effectively, or by a residual frontier that is unlikely to disappear through prompt-level context improvements alone \cite{RGFL2026,ContextBench2026,liu2023_lostinthemiddle}.
This missing decomposition is important because these failure sources imply very different research bets, yet they are often conflated in end-to-end comparisons of repository-level APR systems \cite{ref08_dissecting_leaderboards_2025,ref10_rigorous_eval_agents_2025,RepoFixEval2026}.

In this paper, we study that post-localization gap on \swebenchlite{} with three representative repository-level RAG-APR paradigms: \agentless{}, \kgcompass{}, and \experepair{}.
We design a protocol that isolates several post-localization factors, including Oracle Localization, within-pool \bestofk{} sampling, controlled context augmentation under fixed interfaces, and a unified oracle check \cite{RGFL2026,chen2021_codex,SC,ContextBench2026}.
This lets us ask a narrower but important question:

\vspace{-2mm}
\hypobox{
``Once localization gets stronger, what can still be recovered, which gains stay bounded, and what remains jointly unsolved?''
}
\vspace{-2mm}

Compared with prior work, our goal is not to propose another repair system or another retrieval module.
Instead, we connect localization, search, evidence attribution, and residual-frontier analysis in one controlled empirical study.
Our results show a consistent pattern: stronger localization helps, but it does not remove most of the gap; within-system search and informative added context still recover cases under bounded settings; and prompt-level fusion still leaves a broad jointly-unsolved frontier.
The common-wrapper check also shows that some oracle gains depend on how the final repair interface is built, especially for \agentless{}.

The study is organized around four research questions that separate oracle localization, within-pool search headroom, fixed-interface added context, and the residual frontier; we state them formally in Section~\ref{sec:study-design}.

\vspace{1mm}
This paper makes the following contributions:
\begin{itemize}[leftmargin=*,topsep=2pt]
  \item \textbf{Controlled Post-Localization Study.} We run a four-part controlled study on \swebenchlite{} across three representative repository-level RAG-APR paradigms.
  \item \textbf{Bounded Post-Localization Headroom.} We characterize how much recoverable headroom remains after stronger localization, including the bounded within-system search gains available inside sampled candidate pools.
  \item \textbf{Control-Aware Evidence Attribution.} We test fixed-interface added context with per-condition same-token filler controls, same-repository hard negatives, token-based budget audits, and a common-wrapper oracle check.
  \item \textbf{Residual Frontier Characterization.} We show that prompt-level gains recover only part of native complementarity and still leave a broad jointly-unsolved frontier under the current protocol.
\end{itemize}

%% file: backgroundv2.tex
\section{Background}
\label{sec:background}

\subsection{Repository-Level APR}
Traditional APR often targeted isolated bugs within one function or one file \cite{monperrus2018_automatic_software_repair_bibliography,gazzola2019_automatic_software_repair,goues2012_genprog,thien2013_semfix,kim2013_par,Martinez2014Astor,Martinez2017RealBugs,Mechtaev2016Angelix}.
Repository-level APR is harder because real issues can span multiple files, depend on project-specific structure, and require the model to reason over a codebase that is much larger than one prompt \cite{SWE-bench}.
Even with longer context windows, feeding the whole repository is still impractical and can dilute useful signals \cite{liu2023_lostinthemiddle}.
RAG is therefore a practical way to build a bounded repair context \cite{lewis2020_rag}.
In this setting, the central design problem is not only whether to retrieve evidence, but also what evidence to retrieve and how to organize it before repair.

\subsection{RAG-APR Decomposition}
To bridge the gap between repository-scale code and bounded LLM contexts, prior work has converged on a RAG-based formulation \cite{lewis2020_rag,InferFix,RAP-Gen}.
In this paper, we use \emph{RAG-APR} to refer to the following three-stage paradigm:
\begin{enumerate*}[label=(\arabic*), itemjoin=\quad]
    \item \textbf{Localization:} narrow the search space to candidate files, functions, or lines relevant to the issue.
    \item \textbf{Context Construction:} assemble compact evidence into a bounded prompt.
    \item \textbf{Repair:} generate candidate patches with the LLM conditioned on the constructed context.
\end{enumerate*}

\subsection{Representative RAG-APR Paradigms}
Our study focuses on \agentless{}, \kgcompass{}, and \experepair{} because they share this bounded-context workflow but differ in how they build repair evidence.
\agentless{} uses a staged workflow with hierarchical localization, compressed repository views, and test-based patch selection \cite{Agentless}.
\kgcompass{} builds a repository-aware knowledge graph, ranks function-level candidates with graph proximity, and provides entity paths as structured evidence for prompting \cite{KGCompass}.
\experepair{} uses dual memory to retrieve past demonstrations and semantic insights, then injects them into iterative test and patch generation before final validation \cite{ExpeRepair}.

These methods are therefore not orthogonal end-to-end: they still share localization, bounded prompts, and downstream patch generation logic.
What differs most is the evidence carrier.
\agentless{} relies on staged workflow control and compressed repository structure, \kgcompass{} relies on graph-structured repository relations, and \experepair{} relies on experience-based repair memory.
Recent repository-level repair systems span staged pipelines, structure-aware retrieval, memory-augmented repair, and broader agent workflows \cite{Agentless,KGCompass,ExpeRepair,SWE-Agent,zhang2024_autocoderover,ref20_openhands_2024,ref23_sgagent_2026,ref25_reporepair_2026,yang2025surveyllmbasedautomatedprogram,Tao2025RetrievalAugmentedCodeGenerationSurvey}.
We therefore use these three systems as a compact comparison set for evidence-construction strategies in repository-level bounded-context repair.

%% file: relatedworkv2.tex
\section{Related Work}
\label{sec:related}

Recent repository-level APR work is largely built on the \swebench{} family, which now includes multimodal, multilingual, long-horizon, and live variants \cite{SWE-bench,ref03_swebench_multimodal_2024,ref04_multi_swebench_2025,ref05_swebench_pro_2025,ref06_swebench_goes_live_2025}.
Most system papers still aim at end-to-end gains through staged pipelines, agents, structure-aware search, graph retrieval, or repair memory \cite{Agentless,SWE-Agent,zhang2024_autocoderover,KGCompass,ExpeRepair,yang2025surveyllmbasedautomatedprogram}.
These works define the main design space of repository-level RAG-APR, but they do not directly study what remains hard after localization is strengthened.

Recent empirical analyses ask what current benchmarks and leaderboards actually measure.
Prior audits study leaderboard architectures, benchmark composition, evaluation rigor, solved-issue correctness, memory effects, and benchmark mutation \cite{ref08_dissecting_leaderboards_2025,ref09_whats_in_benchmark_2026,ref10_rigorous_eval_agents_2025,ref11_solved_issues_correctness_2025,ref07_swebench_memory_2025,ref12_saving_swebench_2025}.
RepoFixEval further decomposes repository-level repair into issue discovering, localization, and fixing \cite{RepoFixEval2026}.
This line is closest to our paper, but it still stops short of measuring how much headroom remains after localization is strengthened.

A related line of work studies retrieval quality and context use more directly \cite{Tao2025RetrievalAugmentedCodeGenerationSurvey}.
RepoBugs and its RLCE study show that naive repository context extraction can easily become redundant and imprecise \cite{RLCE}.
Prior retrieval-augmented repair systems explore exemplars, static analysis, templates, and knowledge graphs as evidence carriers \cite{CEDAR,RAP-Gen,InferFix,T-RAP,DSrepair}.
RepoBench, ContextBench, and SWE-ContextBench evaluate context retrieval and use more directly, rather than relying only on final solve rate \cite{zhang2023_repocoder,ContextBench2026,ref16_swe_contextbench_2026}.
Lost in the Middle similarly shows that relevant evidence is not reliably used just because it appears in a long prompt \cite{liu2023_lostinthemiddle}.
These studies motivate our fixed-interface added context probes and our same-token filler and hard-negative controls.

Localization remains a major focus because better fault evidence can improve repair \cite{Zeller2002DeltaDebugging,Abreu2009SBFL,Wotawa2002MBDvsSlicing,wu2023largelanguagemodelsfault,Kang2024LLMExplainableFaultLocalization,Liu2019FLBias}.
RGFL studies this with an oracle-style upper-bound analysis \cite{RGFL2026}.
Pass@k and self-consistency motivate our analysis of within-system search headroom \cite{chen2021_codex,SC,Kang2022PatchPrioritization,Hanna2025RLMutationAPR}.
Our question is narrower: after localization is strengthened on \swebenchlite{}, how much headroom remains, how fast does search saturate, and how much frontier stays jointly unsolved?
Thus, unlike prior system papers or single-factor analyses, our paper contributes a controlled post-localization diagnosis rather than another repair method.

%% file: experimental_setupv2.tex
\section{Approach}
\label{sec:approach}

\subsection{Overview}
\label{sec:approach-overview}

This paper uses controlled interventions to study post-localization headroom under a fixed benchmark split.
We keep the benchmark split and harness fixed, then vary localization, within-system search, and added context while leaving the evaluation pipeline unchanged.
All primary comparisons use \swebenchlite{}.

\input{tables/fig-oracle-intervention}

Figure~\ref{fig:oracle-intervention} shows the shared study flow.
The issue description is the shared input, each system first builds its own candidate repair blocks, Oracle can then inject gold-derived pre-patch spans, each system still builds its own final repair prompt, the model generates a patch, and the harness validates the result.
The later interventions change only selected steps in this flow.
We read their effects through outcome transitions, cross-system overlap, and residual-frontier audits.

\subsection{Oracle Localization}
\label{sec:oracle-localization}

Oracle Localization is the paper's main counterfactual intervention.
At the oracle step in Figure~\ref{fig:oracle-intervention}, we parse the gold diff and extract pre-patch fault spans
$G = \{(\textit{file}, [\textit{start}, \textit{end}])\}$.
Oracle injects repair blocks derived from $G$ into each system's candidate set before that system builds its own final repair prompt.
Only pre-patch file and line information is exposed; patch text and replacement code are never revealed.
This intervention asks how much end-to-end gap stronger candidate localization removes when the downstream prompt construction, patch generation, and validation logic remain system-specific.

\paragraphbold{Ground-truth extraction.}
For each instance, we parse the gold patch, extract its diff hunks, map them to pre-patch file paths and line ranges, and obtain a set of fault spans $G$.
The same extracted spans are used for all systems, but they are merged into different candidate sets and wrapped into different final prompts.

\paragraphbold{Interface-specific merge rules.}
The span extraction is shared, but the merge rule is system-specific and fixed.
This is a deliberate design choice.
Context construction is one of the main differences among repository-level APR systems.
If we instead standardized all three pipelines with one shared post-oracle prompt builder, the absolute scores might rise, but the comparison would collapse the context-construction differences that this study is meant to preserve.
We therefore keep the oracle spans shared while preserving each system's own context-construction logic.
\textbf{\agentless{}.} Oracle files are moved to the front of found files, and oracle line blocks are prepended within found edit locations, so the native repair interface still consumes the same data structure.
\textbf{\kgcompass{}.} Oracle file-line spans are converted into method-like entries and prepended to the existing final locations method list before the native issue/method evidence block is rebuilt.
\textbf{\experepair{}.} The parsed oracle spans inject the bug locations structure with a single oracle group, while the downstream memory-stage interface and final patch instruction remain unchanged.

The intervention is therefore shared at the span level but interface specific at the merge stage.
It combines shared oracle spans with each system's own prompt-construction logic.
We later pair the native-Oracle analysis with a common-wrapper oracle check that replays Oracle under one common repair wrapper, and one variant also uses one common oracle prompt builder.

\subsection{\bestofktext{} and Ideal Selection}
\label{sec:best-of-k}

To estimate how much additional headroom remains after stronger localization, we evaluate an idealized selector over $K=10$ independently sampled patch candidates per instance under Oracle Localization, following multi-sample evaluation and self-consistency ideas \cite{chen2021_codex,SC}.
\textsc{Solved@}K is 1 if any of the $K$ samples resolves the instance and 0 otherwise.
This gives a within-pool upper bound for within-system candidate diversity plus ideal selection, while keeping each system's repair and validation pipeline unchanged.

For intermediate $K{<}10$, we greedily reorder the sampled 10-patch pools by marginal solved gain and derive within-pool \textsc{Solved@}K from that order.

\subsection{Fixed-Interface Added Context Probes and Controls}
\label{sec:evidence-transfer}

Motivated by the complementarity exposed by the first two interventions, we run a follow-up added context attribution study on two fixed repair interfaces, in the same spirit as process-level context studies that separate retrieved context from downstream use \cite{ContextBench2026}.
Both blocks reuse native oracle prompt artifacts upstream, then integrate extra context immediately before the final patch-generation stage.

\paragraphbold{\agentless{} interface.}
We reuse the native \agentless{} Oracle repair prompt as a fixed interface.
The pipeline first follows the original \agentless{} oracle run and keeps that prompt skeleton unchanged.
We then insert extra context before the final output-format instructions and continue with the same downstream \agentless{} repair call.
On \swebenchlite{} with $K=10$ samples and the official harness, we compare three informative added context conditions:
(i) a \emph{KG-augmented context}, which adds only \kgcompass{} oracle-side issue and method evidence reconstructed from its oracle final locations;
(ii) an \emph{Expe-augmented context}, which adds only the pre-assistant conversational context extracted from \experepair{};
and (iii) \textsc{UnionContext}, which combines both evidence channels.
For each informative condition, we then build two own controls:
(iv) an \emph{exact filler control}, which keeps the same wrapper skeleton and section structure but replaces the added evidence with non-informative filler matched to that condition's total prompt input tokens; and
(v) an \emph{irrelevant hard negative}, which keeps the same wrapper skeleton and raw prompt style but swaps in same-repository evidence from a different issue, again matched to that condition's total prompt input tokens.
We also report the native \agentless{} Oracle \bestofk{} result for the same interface.

\paragraphbold{\experepair{} final-step interface.}
We repeat the same idea on the native \experepair{} final-step interface.
This block also starts from native \experepair{} oracle generation conversations and their final-step prompt artifacts.
We keep the native message-thread structure and final patch-format instruction fixed, integrate only extra pre-generation context immediately before the final patch-generation stage, and continue with the same final-step patch call.
This keeps the native oracle artifacts fixed at the same final-step integration point used in the \agentless{} block.
We compare three informative conditions:
(i) an \emph{\agentlessname{}-augmented context}, which inserts the raw \agentless{} oracle repair prompt;
(ii) a \emph{KG-augmented context}, which inserts the raw or reconstructed \kgcompass{} oracle prompt;
and (iii) \textsc{UnionContext}, which inserts both blocks.
For each informative condition, we then build the same two own controls as above:
(iv) an \emph{exact filler control}, which keeps the same message-thread structure but replaces the added evidence with non-informative filler matched to that condition's total prompt input tokens; and
(v) an \emph{irrelevant hard negative}, which keeps the same message-thread structure and raw prompt style but inserts same-repository evidence from a different issue, again matched to that condition's total prompt input tokens.
We also report the native \experepair{} Oracle \bestofk{} result for the same interface.

\subsection{Diagnostic Measures}

We track three diagnostic views after the interventions.
First, we use an outcome taxonomy with three states: \emph{resolved}, \emph{completed-unresolved}, and \emph{no patch submitted}.
This display-level collapse maps empty patch and not submitted status to \emph{no patch submitted}.
Under the \swebenchlite{} harness, execution-failure cases are still submitted but unresolved outcomes, so we keep them inside \emph{completed-unresolved}.
Second, we measure overlap across the three native systems to quantify how much complementarity remains jointly unrecovered after each stage.
Third, for the residual jointly-unsolved set, we run a heuristic qualitative failure audit on \agentless{} oracle-run logs, the most detailed logs available.

\section{Evaluation}
\label{sec:evaluation}

We organize the evaluation around four RQs: what Oracle Localization removes (RQ1), how much extra headroom within-system candidate diversity and ideal selection can recover (RQ2), whether added context gains come from informative context rather than prompt length (RQ3), and how much frontier remains after those prompt-level gains (RQ4).

\subsection{Research Questions}
\label{sec:study-design}

We organize the study around four research-question groups that separate oracle localization, within-pool search headroom, fixed-interface evidence attribution, and the residual frontier:

\vspace{1mm}
\noindent\textbf{RQ1 (Oracle Gain and Failure Concentration):}
\begin{itemize}[leftmargin=*,topsep=2pt]
  \item \textbf{RQ1.1 (Oracle Effect):} How much success does Oracle Localization recover over the native Baseline?
  \item \textbf{RQ1.2 (Wrapper Dependence):} How much of that gain remains under a shared wrapper and a shared oracle builder?
  \item \textbf{RQ1.3 (Failure Concentration):} After localization is strengthened, where does the remaining failure mass concentrate?
\end{itemize}

\noindent\textbf{RQ2 (Selection Headroom and Saturation):}
\begin{itemize}[leftmargin=*,topsep=2pt]
  \item \textbf{RQ2.1 (Search Headroom):} How much within-pool search headroom remains after stronger localization?
  \item \textbf{RQ2.2 (Selector Sensitivity):} How sensitive is the remaining headroom to the selector used to rank candidates?
\end{itemize}

\noindent\textbf{RQ3 (Fixed-Interface Evidence Attribution):}
\begin{itemize}[leftmargin=*,topsep=2pt]
  \item \textbf{RQ3.1 (Context vs Controls):} Do informative added context conditions outperform their own same-token fillers and hard negatives?
  \item \textbf{RQ3.2 (Early Gains and Stability):} Do stronger added context conditions help earlier in the $K$ prefix and increase stable solves?
  \item \textbf{RQ3.3 (Fusion Limits and Wrong Evidence):} Does simple prompt fusion saturate the single-source gains, and what happens when the added evidence is wrong?
\end{itemize}

\noindent\textbf{RQ4 (Residual Cross-System Frontier):}
\begin{itemize}[leftmargin=*,topsep=2pt]
  \item \textbf{RQ4.1 (Fixed-Probe Recovery):} How much new frontier can the best fixed informative probes recover beyond the native union?
  \item \textbf{RQ4.2 (Recovery Ceiling):} How much more frontier is recoverable under looser post hoc probe unions and oracle ceilings?
  \item \textbf{RQ4.3 (Frontier Breadth and Failure Modes):} How broad is the residual frontier, and what failure patterns remain there?
\end{itemize}

\subsection{Experimental Setup}
\label{sec:setup}

\subsubsectionbold{Benchmark Split and Harness}\

\paragraphbold{Benchmark.}
We use the \swebenchlite{} \cite{SWE-bench,swebench_lite_website} test split because it provides a curated 300-instance subset that preserves the broad distribution and difficulty spectrum of the original benchmark while making controlled, full-factorial evaluation across multiple systems and intervention settings computationally practical.
Each instance consists of a repository snapshot, an issue description, and a held-out test suite for validation.

\paragraphbold{Task split.}
We evaluate all systems and intervention settings on the full set of 300 instances without subsampling, so that comparisons are made on exactly matched tasks rather than being affected by sampling variance.
Instances that fail to produce a patch or complete execution are counted as failures.

\paragraphbold{Evaluation harness.}
We adopt the official \swebenchlite{} harness to align with the benchmark-standard evaluation protocol and ensure reproducible, directly comparable results across systems.
An instance is counted as \emph{resolved} only if the submitted patch passes all tests in the harness.

\subsubsectionbold{Systems and Backbone}
\label{sec:systems}

We compare three repository-level APR systems that represent different RAG-APR design choices:
\agentless{} \cite{Agentless}, a phase-structured localization--repair--validation pipeline;
\kgcompass{} \cite{KGCompass}, a structure-aware system grounded in repository knowledge graphs; and
\experepair{} \cite{ExpeRepair}, a memory-augmented repair system.
Full commands, prompts, and run configurations are provided in the accompanying anonymous artifact.
In the artifact, dense retrieval uses jinaai/jina-embeddings-v2-base-code \cite{jinaembeddings28192token}, and lexical retrieval uses BM25 \cite{ref39_bm25_2009} where applicable.

Our primary cross-system comparisons use \textbf{DeepSeek-V3} \cite{deepseekv3} across all three systems.
To check whether the Oracle effect is tied to one backbone, we also run \agentless{} and \kgcompass{} with GPT-4.1 \cite{openai2025gpt41} under the same Baseline and Oracle settings.

\paragraphbold{Baseline.}
Each system first runs in its native form.
Given the issue description and repository snapshot, it performs its own localization and context construction, then generates and validates a patch with its default control flow.
All interventions below are measured against this baseline.

\subsubsectionbold{Metrics and Statistics}\
\label{sec:eval}

\paragraphbold{Primary metrics.}
Our main end-to-end metric is \emph{Success}, the proportion of resolved instances on the 300-instance \swebenchlite{} split.
For the multi-sample analysis in RQ2, we report \textsc{Solved@}K computed from one fixed sampled pool.
For the direct localization audit in RQ1, we report File hit, File Recall@1, and File Recall@5.
For the fixed-interface added context study in RQ3, we report \textsc{Solved@}10 together with mean prompt input tokens as the main prompt-size measure.
We record character counts in the artifact.

\paragraphbold{Secondary metrics.}
We record completion rate as a complementary view of how much failure mass stays outside harness-completed runs.
Here completion rate means harness-completed runs divided by \swebenchlite{}, so empty patch cases stay outside its numerator.
For RQ4, we summarize recovery over the jointly-unsolved frontier as recovered/frontier cases, with percentages reported when useful.

\paragraphbold{Statistics.}
For proportions, we report Wilson 95\% confidence intervals \cite{ref48_wilson_1927,ref49_bcd_2001}.
For paired binary outcomes on the same benchmark split, we use the exact McNemar test \cite{ref47_mcnemar_1947} and report paired risk differences with percentile bootstrap 95\% confidence intervals over instance resampling.
For the small-N repository-level frontier recovery rates in RQ4, we report exact binomial 95\% confidence intervals.
For the greedy within-pool upper-bound curves in RQ2, we draw percentile bootstrap 95\% bands over the shared instance pool.
For RQ3, we apply Holm correction separately to the six informative-versus-own-exact-filler comparisons and to the six informative-versus-own-hard-negative comparisons.
We also report native-interface paired comparisons.

%% file: tables/fig-oracle-intervention.tex
\begin{figure}[h]
  \vspace{-1mm}
  \centering
  \includegraphics[width=\linewidth]{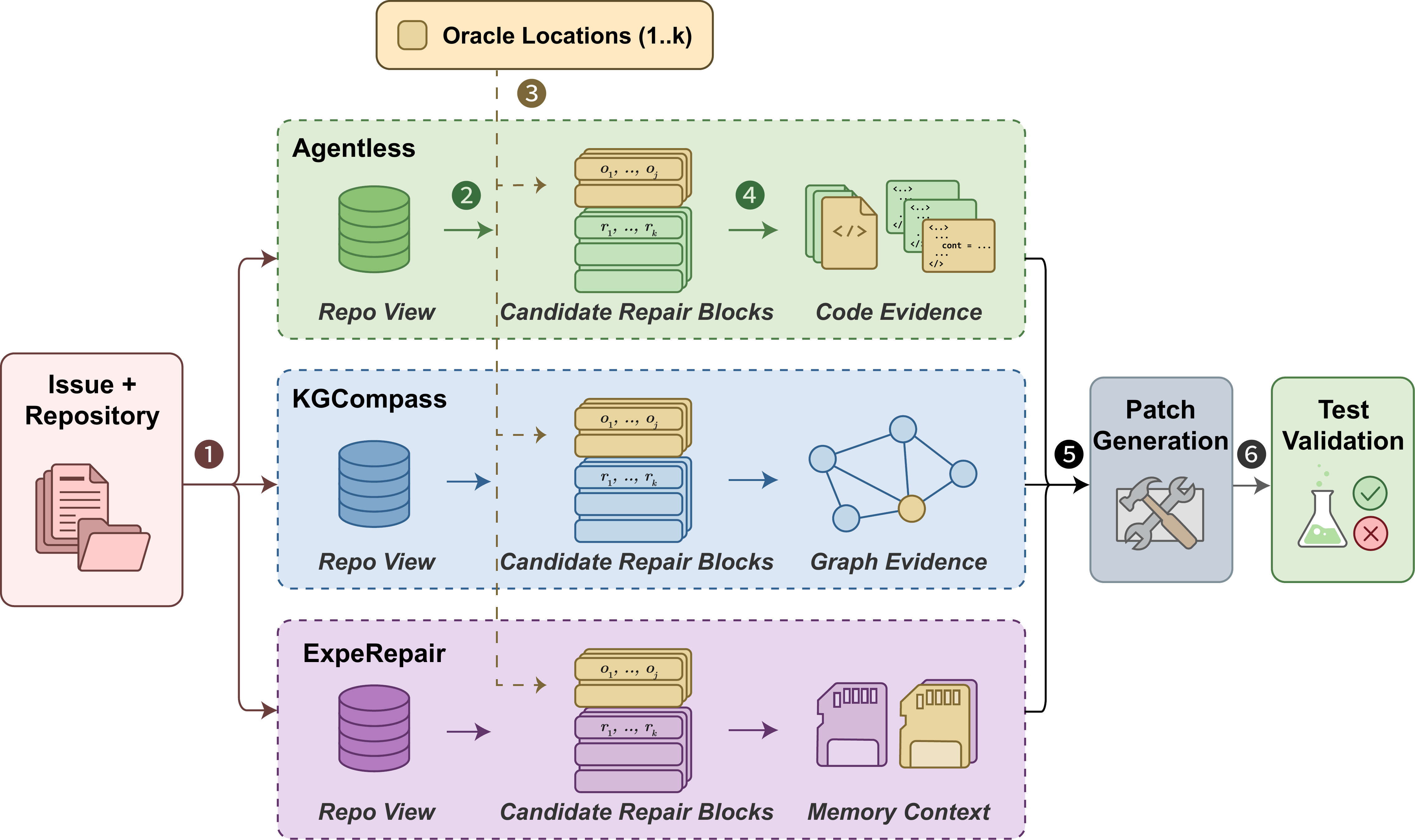}
  \caption{High-level oracle intervention points in the three native pipelines. Oracle injects gold-patch-derived pre-patch file and line spans before each system builds its own final repair prompt.}
  \Description{A six-step pipeline diagram showing shared issue input, native candidate construction, oracle span injection, system-specific prompt building, patch generation, and validation.}
  \label{fig:oracle-intervention}
\end{figure}

%% file: resultsv2.tex
\subsection{RQ1: Oracle Gain and Failure Concentration}
\label{sec:rq1-results}

\subsubsectionbold{Experimental Design}

RQ1 asks how much end-to-end gap Oracle Localization removes once each system keeps its own prompt-construction, patch-generation, and validation flow, and where the remaining failures concentrate.
We compare each system's Baseline run against Oracle Localization on \swebenchlite{} under the native pipelines described in Section~\ref{sec:oracle-localization}.
Because native Oracle keeps each system's own merge rule and prompt builder, RQ1 also includes a common-wrapper check on the shared 300-instance pool.
In that check, `Shared native' and `Shared oracle' hold the repair wrapper fixed, and `Shared-builder oracle' also standardizes the oracle builder.
We report Success, completion rate, paired wins and losses, and a direct localization audit with File hit, File Recall@1, and File Recall@5.

\subsubsectionbold{Experimental Results for RQ1.1 (Oracle Effect)}\

\paragraphtight{Oracle Localization improves all three systems, but success still stays below 50\%.}
Table~\ref{tab:oracle-main} shows that Oracle Localization raises resolved instances from 84 to 121 for \agentless{}, from 88 to 129 for \kgcompass{}, and from 98 to 117 for \experepair{}.
Success here means test-passing under the \swebenchlite{} harness.
Completion rate is completed/total, and confidence intervals are Wilson 95\%.
Completion rate also increases for all three systems, reaching 99.0\%, 98.7\%, and 98.0\% under Oracle.
However, success under Oracle still remains at only 40.3\%, 43.0\%, and 39.0\%.
Oracle raises completion for all three systems, but 61.6\%--79.3\% of the total success gain is associated with higher pass rates among completed runs rather than completion alone.
The remaining gap is therefore large even after the ground-truth fault location is supplied, and most of it remains after completion.

\input{tables/tab-oracle-main}

\paragraphtight{The paired gains are real, but regressions still occur.}
Table~\ref{tab:oracle-paired} reports paired wins and losses on the same instances.
In this paired comparison, wins denote instances where Oracle succeeds and Baseline fails, whereas losses denote the reverse.
All three systems show more wins than losses, with strong evidence for \agentless{} and \kgcompass{}.
The paired risk-difference intervals also stay above zero for all three systems.
At the same time, losses are non-zero for all three systems.
Oracle Localization is therefore a strong intervention, but it is not monotonic.

\input{tables/tab-oracle-paired}

\paragraphtight{GPT-4.1 comparison.}
Under GPT-4.1, Oracle again improves success from 74/300 to 109/300 for \agentless{} and from 55/300 to 126/300 for \kgcompass{}, with paired wins/losses of 40/5 and 78/7, which matches the main-backbone pattern of clear gains and more wins than losses.

\subsubsectionbold{Experimental Results for RQ1.2 (Wrapper Dependence)}\

\paragraphtight{A common-wrapper check shows different system responses under a shared wrapper.}
Table~\ref{tab:oracle-common-wrapper} compares three interface-controlled variants on the shared 300-instance pool.
The asymmetry is sharpest on \agentless{}: under the shared wrapper, success changes only from 35.7\% to 37.0\%, and the large jump appears only under shared-builder oracle (51.0\%).
By contrast, \kgcompass{} and \experepair{} already retain most of their oracle gain under the shared wrapper, reaching 50.3\% and 51.3\% before builder standardization.
This comparison makes the contribution of wrapper and builder choices visible alongside the shared oracle spans.

\input{tables/tab-oracle-common-wrapper}

\subsubsectionbold{Experimental Results for RQ1.3 (Failure Concentration)}\

\paragraphtight{Baseline localization already helps, but it is still far from enough.}
Table~\ref{tab:rq1-localization-audit} summarizes Baseline file-level hit rates, span-hit success, and whether Oracle regressions already had a Baseline span hit.
Baseline localization is already substantial: file hit reaches 65.3\%--85.3\%, file Recall@1 reaches 35.0\%--74.3\%, and file Recall@5 reaches 51.7\%--84.3\%.
Yet even on span-hit instances, Baseline success is still only 34.8\%--45.5\%.
Most Oracle regressions also already had at least one Baseline span hit: all losses for \agentless{} and \experepair{}, and 9/13 losses for \kgcompass{}.
Localization therefore helps, but it does not by itself explain the remaining failures.

\input{tables/tab-rq1-localization-audit}

\paragraphtight{The instance-level transition paths are still highly similar across all three systems.}
Figure~\ref{fig:oracle-outcomes} tracks the same instances from Baseline to Oracle Localization and then to \bestofk{}.
Gray denotes \emph{no patch submitted}, orange denotes \emph{unresolved}, and green denotes \emph{resolved}; execution-failure cases stay inside \emph{unresolved}.
Across all three systems, Oracle mainly shrinks the \emph{no patch} bucket, from 45/28/24 cases to 2/2/5, while execution failures remain rare at 1$\rightarrow$1, 2$\rightarrow$2, and 0$\rightarrow$1.
Table~\ref{tab:rq1-failure-subtypes} makes the stage changes explicit: the first two columns report counts before and after Oracle, and the last three report transitions between failure buckets.
For \agentless{} and \kgcompass{}, \emph{no patch} corresponds to `empty\_patch'; for \experepair{}, it corresponds to `not\_submitted'.
Table~\ref{tab:rq1-failure-subtypes} also shows that most recovered no-patch cases first become \emph{unresolved}, and that Oracle upgrades many already-completed runs to \emph{resolved}.
\bestofk{} then acts mainly on the remaining Oracle-stage \emph{unresolved} pool, converting 33/41/31 such cases to \emph{resolved}.
This stage view again points to the same RQ1 conclusion: after localization improves, the main remaining mass is submitted but unresolved repair attempts.

\input{tables/fig-oracle-outcomes}

\input{tables/tab-rq1-failure-subtypes}

\find{{\bf [RQ-1] Findings:} Oracle Localization closes part of the gap, but success still stays below 50\%. Most recovered gain comes from better pass rates among completed runs rather than completion alone, and the common-wrapper check shows that \agentless{} is much more builder-mediated than \kgcompass{} and \experepair{}. {\bf Insights:}
Stronger localization mainly reduces no-patch failures; the main bottleneck then shifts to post-submission repair.
}

\subsection{RQ2: Selection Headroom and Saturation}
\label{sec:rq2-results}

\subsubsectionbold{Experimental Design}

RQ2 asks whether stronger localization leaves substantial within-system post-localization headroom, or only a bounded amount inside the sampled pools.
Under Oracle Localization, we sample $K=10$ candidate patches per instance and evaluate two views on the same fixed pool.
The main \textsc{Solved@}K upper-bound view greedily reorders the pool by marginal solved gain before taking the top-$K$ prefix.
We then replay fixed prefix orders on that same pool, including the raw sample-index order and simple rerankers, to ask how much of that upper bound is practically reachable.
That raw sample-index order is the system's native generation order rather than a shared temperature schedule.

\subsubsectionbold{Experimental Results for RQ2.1 (Search Headroom)}\

\paragraphtight{\bestofk{} adds meaningful headroom over Oracle, but most of it is already recovered by $K{=}5$.}
Table~\ref{tab:selection-summary} compares Oracle Localization success rate with \textsc{Solved@}5 and \textsc{Solved@}10.
The Oracle Localization rate is taken from Table~\ref{tab:oracle-main}.
\textsc{Solved@}5 and \textsc{Solved@}10 are computed from the same 10-sample pools after greedily reordering samples by marginal solved gain.
\bestofk{} adds 9.7--13.3 points over the Oracle single-run success rates, so candidate diversity and ideal selection still matter.
But the stronger saturation result is that this gain is already almost exhausted by $K{=}5$.
The gain from $K{=}5$ to $K{=}10$ is only 1.3--1.7 points, and the greedy $K{=}5$ prefix already captures 86.2\%--87.5\% of the total Oracle-to-\textsc{Solved@}10 headroom.
Even after greedy reordering, \textsc{Solved@}10 still remains below 60\% for all three native systems.
The remaining search headroom is therefore real but strongly front-loaded.

\input{tables/tab-selection-summary}

\paragraphtight{The greedy upper-bound curves rise early and then flatten.}
Figure~\ref{fig:best-of-k-trends} keeps the greedy within-pool upper bound as the reference line and adds fixed-pool selector replay on the same oracle patch pools.
Across the three panels, the red band marks the bootstrap 95\% interval for that greedy upper bound on the shared instance pool, and the green band marks the span across the fixed random seeds.
The greedy upper-bound curves still rise early and flatten by around $K{=}5$, so the bounded-headroom conclusion remains unchanged.
Read these curves as an early-prefix diagnosis: most recoverable gain is already present in the first few positions of the sampled pool, so later samples add little extra headroom.

\input{tables/fig-best-of-k-trends}

\subsubsectionbold{Experimental Results for RQ2.2 (Selector Sensitivity)}\

\paragraphtight{Selector replay shows that the pool often has a solution, but the early prefix still misses it.}
Figure~\ref{fig:best-of-k-trends} keeps the two most informative non-required references: the raw sample order and cluster-diversity reranking.
Table~\ref{tab:rq2-selector-k5-others} reports the omitted fixed-order and learned rerankers at $K{=}5$.
Those omitted rows do not change the main picture: reverse order helps \agentless{} and \kgcompass{} but hurts \experepair{}, LLM reranking is strongest only on \kgcompass{}, and no omitted selector beats the raw sample order on \experepair{}.
Native sample order therefore already carries system-specific search signal.
At the same time, it still trails the greedy $K{=}5$ upper bound by 3.3 points on \agentless{}, 3.3 on \kgcompass{}, and 1.3 on \experepair{}, so system-native ordering is itself a clear improvement target.
Even if we pick the strongest available selector for each system at $K{=}5$, a 1.3--2.0 point gap to the greedy upper bound still remains.

\input{tables/tab-rq2-selector-k5-others}

\paragraphtight{Taken together, the selector curves point to a two-part bottleneck.}
By $K{=}5$, the greedy upper bound has already captured 86.2\%--87.5\% of the total within-pool headroom, but the default practical reranker still leaves a visible early-prefix gap, especially on \kgcompass{}.
Patch-family clustering suggests why.
The fixed 10-patch pools are not very diverse to begin with: they contain only 1.68--1.83 approximate patch families on average, and only 10.3\%--14.2\% of solvable instances contain more than one solved family.
Yet the default $K{=}5$ prefixes still cover only 1.35--1.43 unique families on average, leaving about 3.5 duplicate slots inside the prefix.
Cluster-diversity reranking pushes family coverage close to full coverage, but exact success still stays 3.4--4.4 points below the family-hit rate.
The bottleneck is therefore not just more samples or a universally better scorer.
Small prefixes must first reach the right patch family, and then rank the right variant early inside that family.

\find{{\bf [RQ-2] Findings:} Extra sampling still improves Oracle, but the gain is bounded and strongly front-loaded: by $K{=}5$, greedy selection already captures 86.2\%--87.5\% of the total available within-pool headroom. The remaining headroom is selector-sensitive, but no single selector closes it consistently, and the default practical reranker still leaves a visible early-prefix gap. {\bf Insights:}
The main bottleneck is not more samples alone. It is reaching the right patch family early and then ranking the right variant within that family.
}

\subsection{RQ3: Fixed-Interface Evidence Attribution}
\label{sec:rq3-results}

\subsubsectionbold{Experimental Design}

Since RQ2 shows that the search within the system still helps but saturates early, RQ3 asks whether additional gains can instead be recovered by adding cross-paradigm context blocks after each interface has already built its native oracle prompt artifacts.
On the \agentless{} side, this is the native oracle repair interface.
On the \experepair{} side, this is the native oracle conversation carried to the final-step patch-generation interface.
In both blocks, the added evidence is integrated only before the final patch-generation stage.
Each informative condition is evaluated against its own same-token filler control and same-repository hard negative, with prompt input tokens as the main budget unit and Holm-corrected paired tests as described in Section~\ref{sec:eval}.

\subsubsectionbold{Experimental Results for RQ3.1 (Context vs Controls)}\

\paragraphtight{Informative evidence beats its controls, but more context is not always better.}
Table~\ref{tab:evidence-transfer} reports each informative condition alongside its own same-token filler control and same-repository hard negative, with prompt input tokens as the main budget unit.
After Holm correction, five of the six informative-versus-filler comparisons and all six informative-versus-hard-negative comparisons remain significant; in the native-interface paired comparisons, only the two lighter single-source transfers are not significant.
Budget matching is tight: the mean prompt-token gap between each informative row and its own filler or hard negative stays below one token on average, although the hard negatives still require truncation on 37.7\%--63.0\% of instances.
Tokens are not monotonic.
On \agentless{}, \kgcompass{}-augmented reaches 60.7\% with 13.46k input tokens, whereas the longer \textsc{UnionContext} row reaches 60.0\% with 15.29k.
On \experepair{}, \textsc{UnionContext} is still the top point estimate, but it improves over \kgcompass{}-augmented by only 1.3 points (59.0\% vs.\ 57.7\%) while adding 1.69k more input tokens.
Added evidence therefore helps because it adds useful structure, not because more context is always better.

\input{tables/tab-evidence-transfer}

\subsubsectionbold{Experimental Results for RQ3.2 (Early Gains and Stability)}\

\paragraphtight{The strongest added context rows help early, but their stability effects differ by wrapper.}
Within the same sampled 10-patch pools, these gains do not appear only at $K{=}10$.
On \agentless{}, \kgcompass{} reaches 48.3\% at $K{=}1$ and 54.0\% at $K{=}2$, already above native $K{=}3$ and $K{=}5$; by $K{=}2$, \textsc{UnionContext} has already realized about 89\% of its final gain at $K{=}10$.
On \experepair{}, \textsc{UnionContext} reaches 45.7\% at $K{=}1$ and 51.7\% at $K{=}2$, and by $K{=}2$ it has already realized essentially all of its final gain at $K{=}10$.
The strongest rows therefore look less like late lucky hits and more like earlier routing into stronger patch families.
The stability pattern, however, is more mixed.
On \agentless{}, \kgcompass{}-augmented reduces all-miss cases by 29 and raises the number of 10/10 stable solves from 63 to 86, whereas \textsc{UnionContext} reduces all-miss by 27 but reaches only 72 stable solves.
On \experepair{}, \textsc{UnionContext} reduces all-miss slightly more than \kgcompass{}-augmented (31 vs.\ 27), but it still ends with fewer 10/10 stable solves (81 vs.\ 86).
The extra source therefore changes the gain shape differently across wrappers: it mostly dilutes stability on \agentless{}, but on \experepair{} it mainly buys earlier coverage with only a small final-ceiling gain.
The lighter transfers behave differently: \agentless{} Expe-only mainly trims full misses but adds many low-frequency solves, whereas \experepair{} \agentless{}-only mostly turns already-solvable cases into 10/10 solves without widening coverage much.

\subsubsectionbold{Experimental Results for RQ3.3 (Fusion Limits and Wrong Evidence)}\

\paragraphtight{The single \textsc{UnionContext} prompt does not dominate the best single-source row.}
On the \agentless{} interface, running the Expe-only and \kgcompass{}-only probes separately solves 196 unique instances, whereas the single \textsc{UnionContext} prompt solves 180.
There are 20 cases that the separate single-source probes recover but the single \textsc{UnionContext} prompt does not, and only 4 cases in the reverse direction.
The same pattern appears on the \experepair{} interface: the separate single-source probes solve 189 unique instances, whereas the single \textsc{UnionContext} prompt solves 177, with a 20-case separate-only advantage versus 8 in the reverse direction.
This fusion limit also appears inside the stronger rows themselves.
On \agentless{}, the longer \textsc{UnionContext} prompt does not improve over the best single-source row, and it converts 26 cases that are 10/10 under \kgcompass{}-augmented into only 1--9/10 solves.
On \experepair{}, \textsc{UnionContext} still improves early coverage over \kgcompass{}-augmented, but it converts 21 of that row's 10/10 cases into 1--9/10 solves while raising final \textsc{Solved@}10 by only 1.3 points.
Simple prompt concatenation therefore does not simply add source-specific gains together; it can also dilute strong single-source trajectories.

\paragraphtight{Wrong evidence is not a neutral control.}
For \agentless{} \textsc{UnionContext}, the same-token hard negative drops from 180 to 63 resolved instances and raises pool-level empty-patch incidence from 41 to 257; for \experepair{} \textsc{UnionContext}, it drops from 177 to 125 and raises the same incidence from 20 to 133.
This drop is too large to read as prompt length alone.
The hard negatives keep the same wrapper skeleton and repository style, but replace task-relevant evidence with the wrong structured constraints.
Added context therefore helps because it supplies the right structured evidence, not because the prompt is merely longer.

\find{{\bf [RQ-3] Findings:} Most informative added context conditions outperform their matched controls, and the largest gains come from \kgcompass{}-based and \textsc{UnionContext} prompts. These gains appear early in the $K$ prefix, but the value of an extra source is not monotonic: on \agentless{}, the best single-source row already matches or exceeds the longer union prompt, while on \experepair{} the union prompt mainly improves earlier coverage and only slightly raises the final ceiling. {\bf Insights:}
Added context helps when it adds task-relevant structure, but simple prompt fusion does not reliably add the gains from multiple sources and can dilute stable single-source gains.
}

\subsection{RQ4: Residual Cross-System Frontier}
\label{sec:rq4-results}

\input{tables/fig-frontier-overlap}

\subsubsectionbold{Experimental Design}

RQ3 shows that informative added context still helps, RQ4 then asks whether these prompt-level gains are enough to explain most of the remaining complementarity, or whether a broader residual frontier still remains.
We answer it with system-overlap progression, a compact frontier-recovery ladder from the native union to progressively looser post hoc ceilings, and a qualitative log audit.
This design uses the overlap and frontier diagnostics introduced in Section~\ref{sec:approach} and reports frontier recovery rates with exact binomial confidence intervals as described in Section~\ref{sec:eval}.

\subsubsectionbold{Experimental Results for RQ4.1 (Fixed-Probe Recovery)}\

\paragraphtight{Prompt-level gains are mostly consolidation gains rather than frontier-opening gains.}
Figure~\ref{fig:frontier-overlap} shows solved-instance overlap across the three native systems under Baseline, Oracle Localization, and \bestofk{}.
Region labels in the figure report the exact overlap counts, and across the three panels the jointly-unsolved set shrinks from 161 to 130 to 101.
Under \bestofk{}, the native three-system union reaches 199/300 solved instances, but 101 instances remain jointly unsolved, so a large frontier still remains.
Figure~\ref{fig:rq4-probe-overlap} then shows only small frontier recovery beyond that native union: +6 for the best single probe, +10 for the best two-probe union, and +14 for the post hoc ceiling.
Table~\ref{tab:rq4-win-frontier-gap} then shows why: most probe-local wins stay inside the native union rather than opening new frontier cases. Across all six probes, only 5.0\%--15.8\% of wrapper-local wins open new frontier cases, whereas 84.2\%--95.0\% stay inside the native union.
Even the stronger rows follow the same pattern.
\kgcompass{} context on \experepair{} and \textsc{UnionContext} on \experepair{} post 38 and 44 wins over their own native wrapper, but only 6 of those wins in each case expand the native union.
\textsc{UnionContext} on \agentless{} is more concentrated still, with 34 of its 36 wins (94.4\%) staying inside the native union.
Even the best fixed single probe therefore recovers only 6 of the 101 native frontier cases (5.9\%, exact 95\% CI 2.2--12.5) and still leaves 95 jointly unsolved.

\input{tables/tab-rq4-win-frontier-gap}

\subsubsectionbold{Experimental Results for RQ4.2 (Recovery Ceiling)}\

\paragraphtight{Even looser post hoc ceilings remain small.}
The full ladder in Figure~\ref{fig:rq4-probe-overlap} stays shallow: the best three-probe union reaches 212/300 (+13), the best four-probe union reaches 213/300 (+14), and the all-6 oracle stays at the same 213/300.
Even under that favorable ceiling, recovery rises only to 14 of the 101 frontier cases (13.9\%, exact 95\% CI 7.8--22.2), leaving 87 residual cases.
The last two probes therefore add no extra frontier recovery beyond the best four-probe union.
This low ceiling is also not just a fixed-probe artifact: even allowing repo-adaptive choice among single informative probes would still recover only 10 of the 101 frontier cases, which is the same order as the best two-probe union.

\input{tables/fig-rq4-probe-overlap}

\subsubsectionbold{Experimental Results for RQ4.3 (Frontier Breadth and Failure Modes)}\

\paragraphtight{The residual frontier is concentrated, broad, and still hard at the repository level.}
The native \bestofk{} jointly-unsolved frontier spans 11 of the 12 repositories in \swebenchlite{}, but it is not uniformly spread.
\texttt{sympy} and \texttt{django} alone account for 65 of the 101 frontier cases.
Even there, recovery stays small: the best fixed single probe recovers only 4 of those 65 cases, and even a repo-wise oracle over single probes recovers only 5.
At the same time, five frontier-bearing repositories remain zero-gain even under that favorable repo-wise oracle, covering 13 frontier cases in total.
The residual frontier is therefore both concentrated and broad: a few large repositories contribute most of the raw frontier mass, but unrecovered cases still span almost the whole benchmark.

\paragraphtight{Qualitative log-audit patterns in the residual frontier.}
Figure~\ref{fig:hard-core-failures} summarizes a qualitative audit of the original 101 jointly-unsolved native cases.
The labels come from heuristic annotation of \agentless{} oracle-run evaluation logs.
Within that scope, the most common labels are wrong API or runtime exceptions, incomplete fixes, and missing imports or \texttt{NameError}s.
These patterns are not only common but also hard to recover under the current probe family: wrong API, incomplete fix, and assertion mismatch together account for 60 of the 101 frontier cases, yet the all-6 oracle recovers only 4 of them.
The residual frontier is therefore not just broad across repositories; it also remains stubborn within the dominant post-localization failure modes.

\input{tables/fig-hard-core-failures}

\find{{\bf [RQ-4] Findings:} Prompt-level fusion recovers only a small part of the remaining complementarity. Across all six informative probes, most wrapper-local wins stay inside the native union rather than opening new frontier cases, and even the favorable post hoc ceiling recovers only 14 of the 101 native frontier cases. The residual frontier is both concentrated and broad: a few large repositories contribute most of the raw frontier mass, but unrecovered cases still span almost the whole benchmark. {\bf Insights:}
Most added context gains reinforce existing strengths rather than unlock new frontier, and the remaining frontier persists both across repositories and within the dominant post-localization failure modes.
}

%% file: tables/tab-oracle-main.tex
\begin{table}[h]
  % \vspace{-2mm}
  \centering
  \small
  \caption{Baseline and Oracle Localization on \swebenchlite{}. Success is resolved/300, completion rate is completed/300, and success among completed is resolved/completed.}
  \label{tab:oracle-main}
  \resizebox{\linewidth}{!}{
  \begin{tabular}{l l r r r}
    \toprule
    System & Setting & Success (95\% Wilson CI) & Completion rate & Success among completed \\
    \midrule
    \multirow{2}{*}{\agentless{}} & Baseline & 28.0\% (23.2--33.3) & 84.7\% & 33.1\% \\
     & Oracle & \textbf{40.3\%} (34.9--46.0) & 99.0\% & 40.7\% \\
    \midrule
    \multirow{2}{*}{\kgcompass{}} & Baseline & 29.3\% (24.5--34.7) & 90.0\% & 32.6\% \\
     & Oracle & \textbf{43.0\%} (37.5--48.7) & 98.7\% & 43.6\% \\
    \midrule
    \multirow{2}{*}{\experepair{}} & Baseline & 32.7\% (27.6--38.2) & 92.0\% & 35.5\% \\
     & Oracle & \textbf{39.0\%} (33.7--44.6) & 98.0\% & 39.8\% \\
    \bottomrule
  \end{tabular}
  }
  \vspace{-2mm}
\end{table}

%% file: tables/tab-oracle-paired.tex
\begin{table}[h]
  % \vspace{-2mm}
  \centering
  \small
  \caption{Paired success changes from Baseline to Oracle Localization on \swebenchlite{}. Wins mean Oracle-only successes, and losses mean Baseline-only successes.}
  \label{tab:oracle-paired}
  \resizebox{\linewidth}{!}{
  \begin{tabular}{l r r r}
    \toprule
    System & \shortstack{Paired success\\change (points, 95\% CI)} & Wins / losses & McNemar $p$ \\
    \midrule
    \agentless{} & +12.3 (7.7--17.0) & \textbf{46 / 9}  & $4.34\times10^{-7}$ \\
    \kgcompass{} & +13.7 (8.3--18.7) & \textbf{54 / 13} & $4.47\times10^{-7}$ \\
    \experepair{} & +6.3 (1.0--11.7)  & \textbf{43 / 24} & $2.71\times10^{-2}$ \\
    \bottomrule
  \end{tabular}
  }
  \vspace{-2mm}
\end{table}

%% file: tables/tab-oracle-common-wrapper.tex
\begin{table}[ht]
  \vspace{1mm}
  \centering
  \small
  \caption{Common-wrapper oracle check on the shared 300-instance pool. Shared native keeps one shared repair wrapper, shared oracle adds oracle spans under that wrapper, and shared-builder oracle also shares the oracle prompt builder.}
  \label{tab:oracle-common-wrapper}
  \resizebox{\linewidth}{!}{
  \begin{tabular}{l r r r r r r}
    \toprule
    System & \shortstack{Shared\\native} & \shortstack{Shared\\oracle} & \shortstack{Shared-builder\\oracle} & \shortstack{$\Delta$ oracle\\vs shared native} & \shortstack{$\Delta$ builder\\vs shared native} & \shortstack{$\Delta$ oracle\\vs shared builder} \\
    \midrule
    \agentless{} & 35.7\% & 37.0\% & 51.0\% & +1.3 & +15.3 & \textbf{-14.0} \\
    \kgcompass{} & 17.0\% & 50.3\% & 51.0\% & +33.3 & +34.0 & -0.7 \\
    \experepair{} & 40.3\% & 51.3\% & 51.0\% & +11.0 & +10.7 & +0.3 \\
    \bottomrule
  \end{tabular}
  }
  % \vspace{-2mm}
\end{table}

%% file: tables/tab-rq1-localization-audit.tex
\begin{table}[h]
  % \vspace{-2mm}
  \centering
  \small
  \caption{RQ1 localization audit from the Baseline runs. Gold file and gold span mean overlap with the gold patch location.}
  \label{tab:rq1-localization-audit}
  \resizebox{\linewidth}{!}{
  \begin{tabular}{l c c c c c c c}
    \toprule
    System & \shortstack{Gold file\\in set} & \shortstack{Gold file\\top-1} & \shortstack{Gold file\\top-5} & \shortstack{Success with\\gold span} & \shortstack{Success with\\no gold span} & \shortstack{Oracle losses\\with top-1\\gold span} & \shortstack{Oracle losses\\with any\\gold span} \\
    \midrule
    \agentless{} & 82.3\% & 71.7\% & 82.3\% & \textbf{38.1\%} & 1.2\% & 9/9 & 9/9 \\
    \kgcompass{} & 65.3\% & 35.0\% & 51.7\% & \textbf{45.5\%} & 17.1\% & 4/13 & 9/13 \\
    \experepair{} & 85.3\% & 74.3\% & 84.3\% & \textbf{34.8\%} & 0.0\% & 24/24 & 24/24 \\
    \bottomrule
  \end{tabular}
  }
  % \vspace{-2mm}
\end{table}

%% file: tables/fig-oracle-outcomes.tex
\begin{figure*}[h]
  \centering
  \begin{minipage}[t]{0.28\textwidth}
    \centering
    \includegraphics[width=\linewidth]{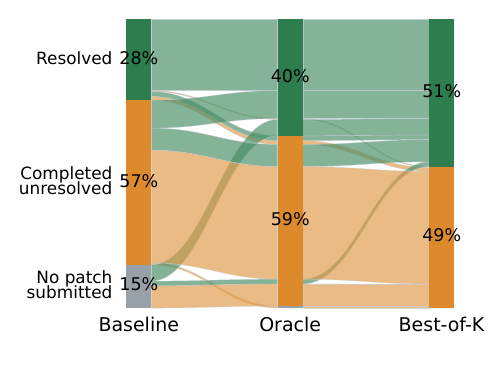}

    {\small (a) \agentless{}}
  \end{minipage}
  % \hfill
  \begin{minipage}[t]{0.28\textwidth}
    \centering
    \includegraphics[width=\linewidth]{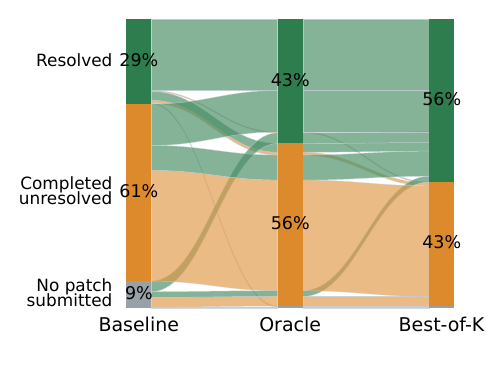}

    {\small (b) \kgcompass{}}
  \end{minipage}
  % \hfill
  \begin{minipage}[t]{0.28\textwidth}
    \centering
    \includegraphics[width=\linewidth]{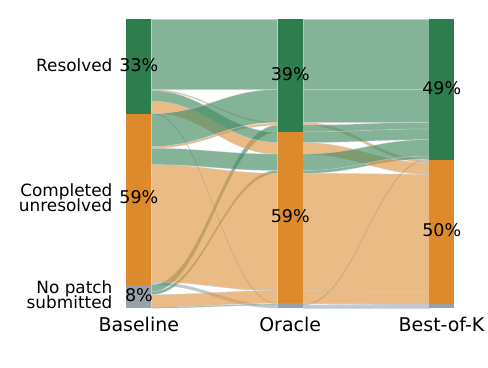}

    {\small (c) \experepair{}}
  \end{minipage}
  \caption{Instance-level three-state outcome transition paths from Baseline to Oracle Localization and then to \bestofk{} ($K=10$). Each ribbon follows the same benchmark instances across stages, with green for resolved, orange for completed but unresolved, and gray for no patch submitted.}
  \Description{Three side-by-side ribbon flow diagrams, one for each system, linking the same instances across three stages.}
  \label{fig:oracle-outcomes}
  \vspace{-2mm}
\end{figure*}

%% file: tables/tab-rq1-failure-subtypes.tex
\begin{table}[h]
  % \vspace{-2mm}
  \centering
  \small
  \caption{RQ1 failure-subtype audit behind Figure~\ref{fig:oracle-outcomes}. The table counts how cases move between the three states in that figure.}
  \label{tab:rq1-failure-subtypes}
  \resizebox{\linewidth}{!}{
  \begin{tabular}{l c c c c c}
    \toprule
    System & \shortstack{No patch\\(Baseline$\rightarrow$Oracle)} & \shortstack{Execution error\\(Baseline$\rightarrow$Oracle)} & \shortstack{No patch\\$\rightarrow$ Unresolved} & \shortstack{No patch\\$\rightarrow$ Resolved} & \shortstack{Unresolved\\$\rightarrow$ Resolved} \\
    \midrule
    \agentless{} & 45$\rightarrow$2 & 1$\rightarrow$1 & 28 & 17 & 29 \\
    \kgcompass{} & 28$\rightarrow$2 & 2$\rightarrow$2 & 16 & 11 & 43 \\
    \experepair{} & 24$\rightarrow$5 & 0$\rightarrow$1 & 15 & 7 & 36 \\
    \bottomrule
  \end{tabular}
  }
  % \vspace{-2mm}
\end{table}

%% file: tables/tab-selection-summary.tex
\begin{table}[h]
  \vspace{-2mm}
  \centering
  \small
  \caption{Selection headroom under stronger localization. Oracle is the single run oracle result, and \textsc{Solved@}5/\textsc{Solved@}10 are within-pool upper bounds from the same 10-sample patch pools.}
  \label{tab:selection-summary}
  \resizebox{\linewidth}{!}{
  \begin{tabular}{l r r r r}
    \toprule
    System & \shortstack{Oracle} & \textsc{Solved@}5 & \textsc{Solved@}10 & \shortstack{Extra gain ($5 \rightarrow 10$)} \\
    \midrule
    \agentless{} & 40.3\% & 49.7\% & 51.0\% & \textbf{+1.3} \\
    \kgcompass{} & 43.0\% & 54.7\% & 56.3\% & \textbf{+1.7} \\
    \experepair{} & 39.0\% & 47.3\% & 48.7\% & \textbf{+1.3} \\
    \bottomrule
  \end{tabular}
  }
\vspace{-2mm}
\end{table}

%% file: tables/fig-best-of-k-trends.tex
\begin{figure*}[h]
  \centering
  \begin{minipage}[t]{0.28\textwidth}
    \centering
    \includegraphics[width=\linewidth]{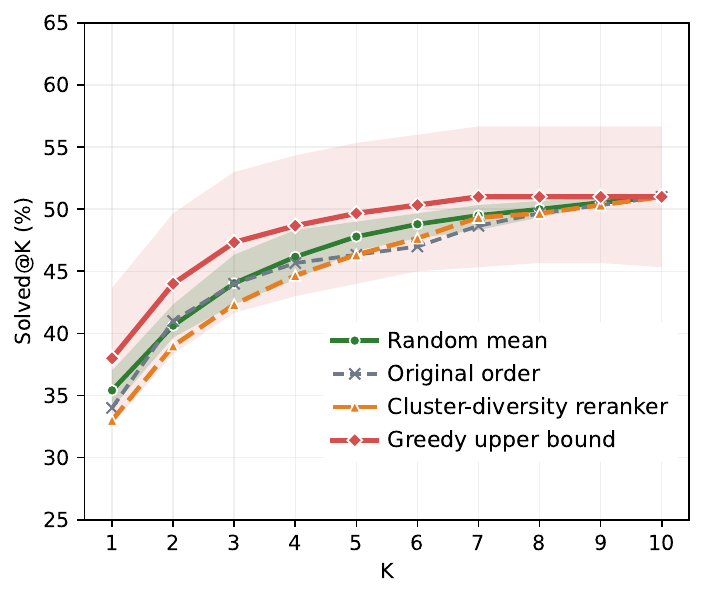}

    {\small (a) \agentless{}}
  \end{minipage}
  % \hfill
  \begin{minipage}[t]{0.28\textwidth}
    \centering
    \includegraphics[width=\linewidth]{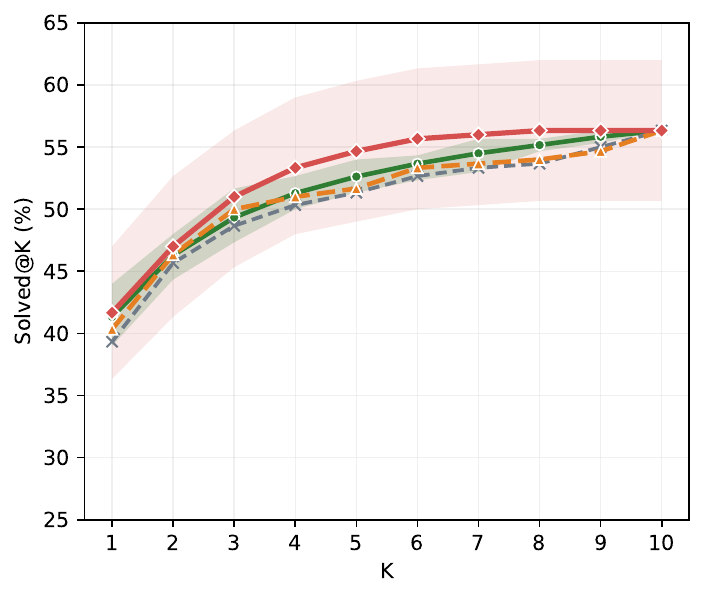}

    {\small (b) \kgcompass{}}
  \end{minipage}
  % \hfill
  \begin{minipage}[t]{0.28\textwidth}
    \centering
    \includegraphics[width=\linewidth]{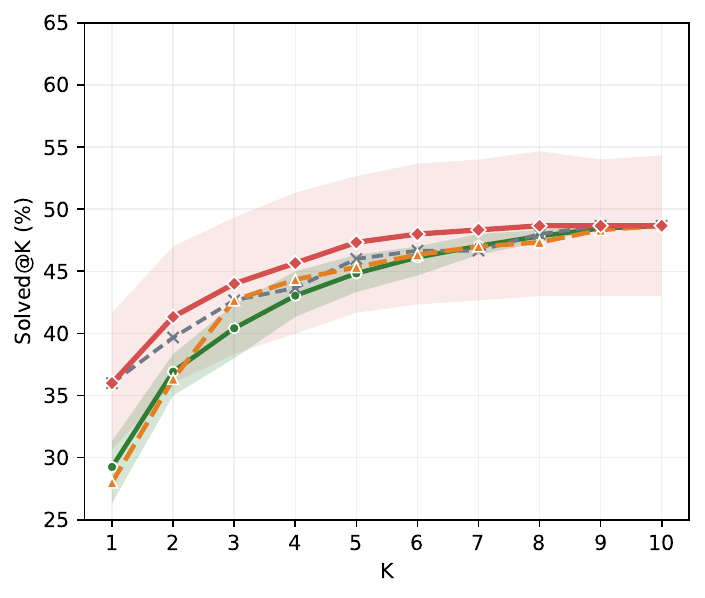}

    {\small (c) \experepair{}}
  \end{minipage}
  \caption{Greedy within-pool upper bounds and fixed-pool selector replay under Oracle Localization for the three native systems. The panels keep the four main references: greedy upper bound, random mean, original order, and cluster-diversity reranking. The red curve and band show the greedy upper bound with a bootstrap 95\% interval, and the green band shows the range across fixed random seeds. Table~\ref{tab:rq2-selector-k5-others} reports the remaining selector rows at $K{=}5$.}
  \Description{Three side-by-side line charts, one for each system, with four strategy traces plus a red reference band and a green range band.}
  \label{fig:best-of-k-trends}
  \vspace{-2mm}
\end{figure*}

%% file: tables/tab-rq2-selector-k5-others.tex
\begin{table}[h]
% \vspace{-2mm}
  \centering
  \small
  \caption{Additional fixed-order and learned selector replay at $K{=}5$ on the same oracle patch pools. These rows are omitted from Figure~\ref{fig:best-of-k-trends} to keep the main plot focused on raw-order signal and family coverage.}
  \label{tab:rq2-selector-k5-others}
\resizebox{\linewidth}{!}{
  \begin{tabular}{l r r r}
    \toprule
    Selector ($K{=}5$) & \agentless{} & \kgcompass{} & \experepair{} \\
    \midrule
    Reverse order & 48.0\% & 52.7\% & 41.0\% \\
    LLM reranker & 46.3\% & 52.7\% & 44.0\% \\
    Cross-encoder reranker & 44.7\% & 48.3\% & 43.3\% \\
    \bottomrule
  \end{tabular}
  }
  % \vspace{-2mm}
\end{table}

%% file: tables/tab-evidence-transfer.tex
\begin{table}[h]
  \centering
  % \vspace{-2mm}
  \caption{Fixed-interface informative probes with their own per-condition controls under Oracle Localization with \bestofk{} ($K=10$). Native, added context, and control columns report \textsc{Solved@}10; tokens are mean input tokens in thousands, and the last column is the share of hard negatives that were truncated to fit the budget.}
  \label{tab:evidence-transfer}
  \resizebox{\linewidth}{!}{%
  \setlength{\tabcolsep}{4pt}
  \begin{tabular}{l l r r r r r r}
    \toprule
    Interface & Added context & \begin{tabular}{@{}c@{}}Native\\prompt\end{tabular} & \begin{tabular}{@{}c@{}}Added\\context\end{tabular} & \begin{tabular}{@{}c@{}}Token-matched\\filler\end{tabular} & \begin{tabular}{@{}c@{}}Same-repo\\hard negative\end{tabular} & \begin{tabular}{@{}c@{}}Input\\tokens (k)\end{tabular} & \begin{tabular}{@{}c@{}}Hard negative\\truncated\end{tabular} \\
    \midrule
     & \begin{tabular}{@{}l@{}}Expe-augmented \\ (\experepairname{} only)\end{tabular} & 51.0\% & 54.7\% & 51.3\% & 47.3\% & 3.47 & 42.0\% \\
    \agentlessname{} & \begin{tabular}{@{}l@{}}\kgcompass{}-augmented \\ (\kgcompass{} only)\end{tabular} & 51.0\% & \textbf{60.7\%} & 51.0\% & 49.3\% & 13.46 & 37.7\% \\
     & \begin{tabular}{@{}l@{}}\textsc{UnionContext} \\ (\kgcompass{} + \experepairname{})\end{tabular} & 51.0\% & \textbf{60.0\%} & 50.0\% & 21.0\% & 15.29 & 55.3\% \\
    \midrule
     & \begin{tabular}{@{}l@{}}\agentlessname{}-augmented \\ (\agentlessname{} only)\end{tabular} & 48.7\% & 52.3\% & 44.7\% & 44.3\% & 3.50 & 50.7\% \\
    \experepairname{} & \begin{tabular}{@{}l@{}}\kgcompass{}-augmented \\ (\kgcompass{} only)\end{tabular} & 48.7\% & \textbf{57.7\%} & 46.0\% & 43.3\% & 4.93 & 43.0\% \\
     & \begin{tabular}{@{}l@{}}\textsc{UnionContext} \\ (\agentlessname{} + \kgcompass{})\end{tabular} & 48.7\% & \textbf{59.0\%} & 45.0\% & 41.7\% & 6.62 & 63.0\% \\
    \bottomrule
  \end{tabular}
  }
  \vspace{-2mm}
\end{table}

%% file: tables/fig-frontier-overlap.tex
\begin{figure*}[h]
  % \vspace{-1mm}
  \centering
  \begin{minipage}[t]{0.28\linewidth}
    \centering
    \includegraphics[width=\linewidth]{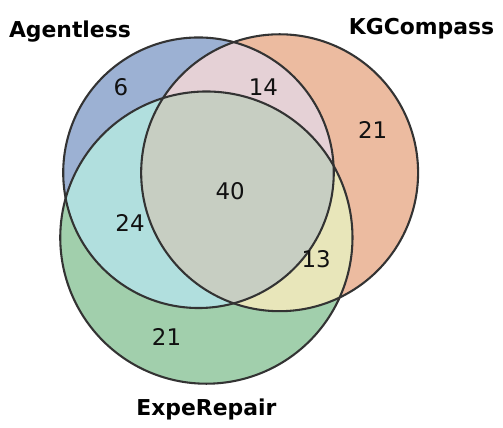}

    \vspace{0.25em}
    \parbox[t]{0.96\linewidth}{\centering \small \textbf{(a)} Baseline. Jointly unsolved = 161; solved by at least one system = 139/300.}
  \end{minipage}
  % \hfill
  \begin{minipage}[t]{0.28\linewidth}
    \centering
    \includegraphics[width=\linewidth]{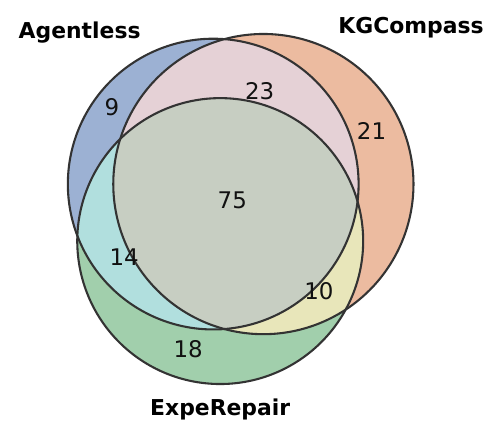}

    \vspace{0.25em}
    \parbox[t]{0.96\linewidth}{\centering \small \textbf{(b)} Oracle Localization. Jointly unsolved = 130; solved by at least one system = 170/300.}
  \end{minipage}
  % \hfill
  \begin{minipage}[t]{0.28\linewidth}
    \centering
    \includegraphics[width=\linewidth]{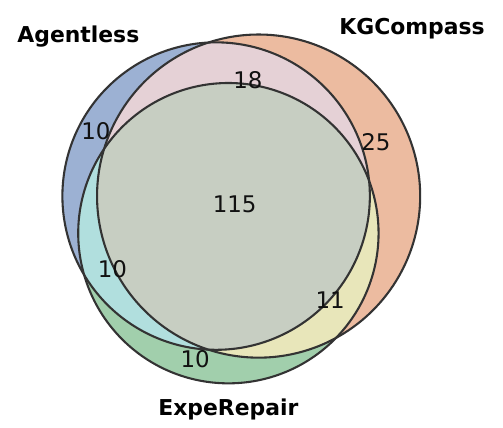}

    \vspace{0.25em}
    \parbox[t]{0.96\linewidth}{\centering \small \textbf{(c)} \bestofk{} ($K=10$). Jointly unsolved = 101; solved by at least one system = 199/300.}
  \end{minipage}
  \caption{Solved-instance overlap across \agentless{}, \kgcompass{}, and \experepair{} under Baseline, Oracle Localization, and \bestofk{}. Each Venn region label is an exact instance count.}
  \Description{Three Venn-style diagrams showing solved-instance overlap across \agentlessname{}, \kgcompassname{}, and \experepairname{} under Baseline, Oracle Localization, and \bestofktext{}, with exact overlap counts in each region.}
  \label{fig:frontier-overlap}
  % \vspace{-2mm}
\end{figure*}

%% file: tables/tab-rq4-win-frontier-gap.tex
\begin{table}[h]
  \centering
  \vspace{-1mm}
  \small
  \caption{Decomposing each informative probe's wins over its own native wrapper under \bestofk{} ($K=10$). The last two columns split those wins into cases already solved by another native system and cases that newly expand the three-system union.}
  \label{tab:rq4-win-frontier-gap}
  \resizebox{\linewidth}{!}{%
  \begin{tabular}{l l r r r}
    \toprule
    Target wrapper & Added context & \begin{tabular}{@{}c@{}}Wins vs\\own native\end{tabular} & \begin{tabular}{@{}c@{}}Already solved\\by native union\end{tabular} & \begin{tabular}{@{}c@{}}New frontier\\cases\end{tabular} \\
    \midrule
    \multirow{3}{*}{\agentlessname{}} & \kgcompass{} & 37 & 32 & 5 \\
     & \experepairname{} & 21 & 18 & 3 \\
     & \textsc{UnionContext} & 36 & 34 & 2 \\
    \midrule
    \multirow{3}{*}{\experepairname{}} & \kgcompass{} & 38 & 32 & 6 \\
     & \agentlessname{} & 20 & 19 & 1 \\
     & \textsc{UnionContext} & 44 & 38 & 6 \\
    \bottomrule
  \end{tabular}
  }
  \vspace{-3mm}
\end{table}

%% file: tables/fig-rq4-probe-overlap.tex
\begin{figure}[h]
  \centering
  % \vspace{-2mm}
  \includegraphics[width=0.7\linewidth]{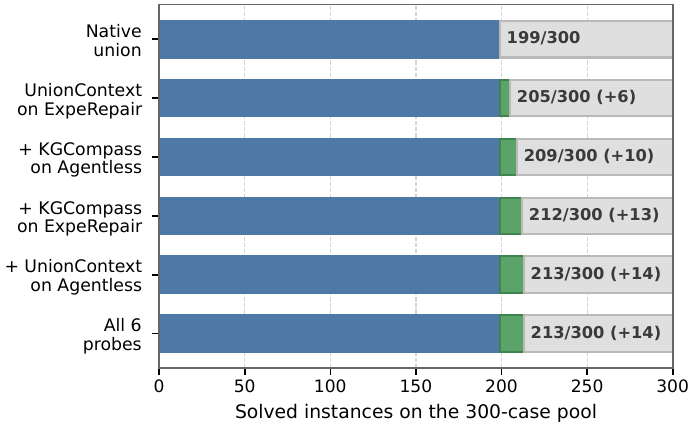}
  \vspace{0.5mm}
  \caption{Compact RQ4 frontier ladder on the shared 300-instance pool. Blue is the native three-system union, green is extra frontier recovered beyond that union, and gray is the remaining frontier.}
  \Description{A six-row horizontal bar chart showing the native union, added frontier gains, and remaining frontier on the shared 300-instance pool.}
  \label{fig:rq4-probe-overlap}
  % \vspace{-2mm}
\end{figure}

%% file: tables/fig-hard-core-failures.tex
\begin{figure}[h]
  \centering
  % \vspace{-2mm}
  \includegraphics[width=0.7\linewidth]{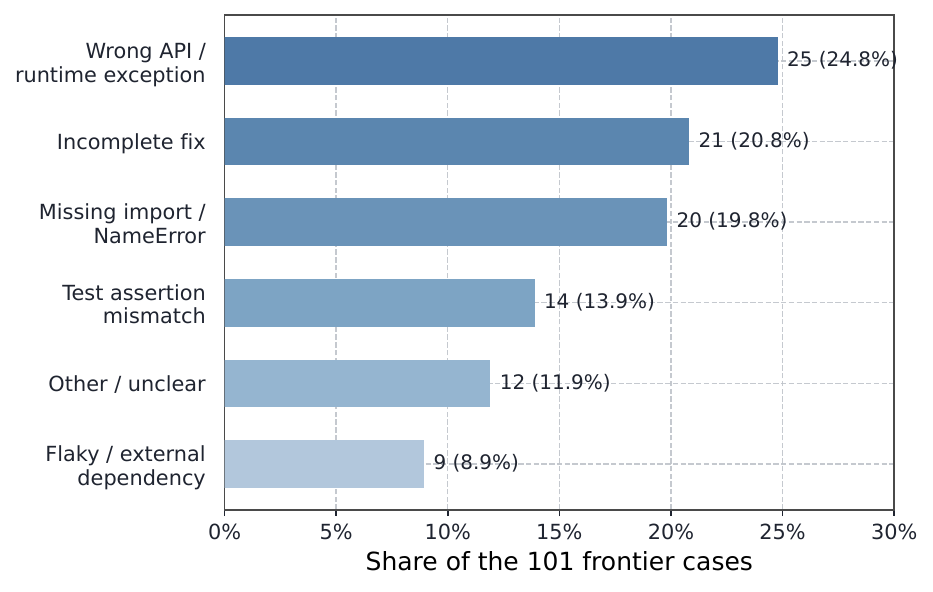}
  \vspace{0.5mm}
  \caption{Qualitative log-audit labels for the 101 jointly-unsolved native cases under \bestofk{}. Bar-end labels show count and share of the 101-case frontier.}
  \Description{A horizontal bar chart of failure categories, with each bar labeled by a count and percentage.}
  \label{fig:hard-core-failures}
  % \vspace{-2mm}
\end{figure}

%% file: analysisv2.tex
\section{Threats to Validity}
\label{sec:threats-validity}

\paragraphbold{Internal.}
The native Oracle intervention keeps each system's own wrapper and prompt builder, so the measured gain is not a pure localization-only effect.
We address this threat with common-wrapper and common-builder checks that separate span-level sharing from interface-level choices.
We therefore interpret the Oracle results as controlled diagnostics under the current protocol, rather than as universal localization-only estimates.

\paragraphbold{Construct.}
Our main outcome is harness-defined test-passing success on \swebenchlite{}.
This is the benchmark's standard outcome, but it does not guarantee full semantic correctness.
We therefore complement it with completion rate, paired wins and losses, localization audits, selector replay, and residual-frontier analysis, and interpret our findings as evidence about benchmark-defined repair behavior.

\paragraphbold{External.}
We evaluate on the full \swebenchlite{} split, a widely used benchmark that covers diverse repositories and supports direct horizontal comparison with prior work \cite{swebench_leaderboards,SWE-bench,swebench_lite_website}.
We also compare three representative RAG-APR paradigms under the same evaluation setup.
However, we do not claim that the same effect sizes will hold for all APR benchmarks, repair interfaces, or model backbones.

%% file: conclusionv2.tex
\section{Conclusion}
\label{sec:conclusion}
In this paper, we study repository-level RAG-APR as a controlled empirical analysis after stronger localization rather than as a new-system competition.
Across the four RQs, stronger localization helps but does not remove most of the remaining gap.
Within-system search and informative added context still recover extra cases under bounded settings, and informative content outperforms matched controls more reliably than a prompt-length-only explanation would predict.
The common-wrapper check also shows different system responses under a shared wrapper, especially for \agentless{}.
Prompt-level fusion still recovers only part of native complementarity, and a large residual frontier remains under the current protocol.
These results argue for a broader post-localization agenda that includes search, evidence quality, and interface design.

\section{Data Availability Statement}
All code and data for this study are available in an anonymized, DOI-minted repository \cite{sourcecode}.
The repository contains the materials needed to reproduce the reported results.
To preserve double-anonymous review, repository metadata are anonymized and will be restored after acceptance.

%% file: main.bib
@INPROCEEDINGS{CEDAR,
  author={Nashid, Noor and Sintaha, Mifta and Mesbah, Ali},
  booktitle={2023 IEEE/ACM 45th International Conference on Software Engineering (ICSE)}, 
  title={Retrieval-Based Prompt Selection for Code-Related Few-Shot Learning}, 
  year={2023},
  volume={},
  number={},
  pages={2450-2462},
  doi={10.1109/ICSE48619.2023.00205}}

@article{SC,
  title={Self-consistency improves chain of thought reasoning in language models},
  author={Wang, Xuezhi and Wei, Jason and Schuurmans, Dale and Le, Quoc and Chi, Ed and Narang, Sharan and Chowdhery, Aakanksha and Zhou, Denny},
  journal={arXiv preprint arXiv:2203.11171},
  year={2022}
}

@inproceedings{RAP-Gen,
author = {Wang, Weishi and Wang, Yue and Joty, Shafiq and Hoi, Steven C.H.},
title = {RAP-Gen: Retrieval-Augmented Patch Generation with CodeT5 for Automatic Program Repair},
year = {2023},
isbn = {9798400703270},
publisher = {Association for Computing Machinery},
address = {New York, NY, USA},
url = {https://doi.org/10.1145/3611643.3616256},
doi = {10.1145/3611643.3616256},
booktitle = {Proceedings of the 31st ACM Joint European Software Engineering Conference and Symposium on the Foundations of Software Engineering},
pages = {146–158},
numpages = {13},
location = {San Francisco, CA, USA},
series = {ESEC/FSE 2023}
}

@inproceedings{InferFix,
author = {Jin, Matthew and Shahriar, Syed and Tufano, Michele and Shi, Xin and Lu, Shuai and Sundaresan, Neel and Svyatkovskiy, Alexey},
title = {InferFix: End-to-End Program Repair with LLMs},
year = {2023},
isbn = {9798400703270},
publisher = {Association for Computing Machinery},
address = {New York, NY, USA},
url = {https://doi.org/10.1145/3611643.3613892},
doi = {10.1145/3611643.3613892},
booktitle = {Proceedings of the 31st ACM Joint European Software Engineering Conference and Symposium on the Foundations of Software Engineering},
pages = {1646–1656},
numpages = {11},
location = {San Francisco, CA, USA},
series = {ESEC/FSE 2023}
}

@INPROCEEDINGS {RLCE,
author = { Chen, Yuxiao and Wu, Jingzheng and Ling, Xiang and Li, Changjiang and Rui, Zhiqing and Luo, Tianyue and Wu, Yanjun },
booktitle = { 2024 IEEE/ACM 46th International Conference on Software Engineering: Companion Proceedings (ICSE-Companion) },
title = {{ When Large Language Models Confront Repository-Level Automatic Program Repair: How Well They Done? }},
year = {2024},
volume = {},
ISSN = {},
pages = {459-471},
doi = {10.1145/3639478.3647633},
url = {https://doi.ieeecomputersociety.org/10.1145/3639478.3647633},
publisher = {IEEE Computer Society},
address = {Los Alamitos, CA, USA},
month =apr}

@inproceedings{T-RAP,
author = {Liu, Pei and Lin, Bo and Qin, Yihao and Weng, Cheng and Chen, Liqian},
title = {T-RAP: A Template-guided Retrieval-Augmented Vulnerability Patch Generation Approach},
year = {2024},
isbn = {9798400707056},
publisher = {Association for Computing Machinery},
address = {New York, NY, USA},
url = {https://doi.org/10.1145/3671016.3672506},
doi = {10.1145/3671016.3672506},
booktitle = {Proceedings of the 15th Asia-Pacific Symposium on Internetware},
pages = {105–114},
numpages = {10},
location = {Macau, China},
series = {Internetware '24}
}

@INPROCEEDINGS {DSrepair,
author = { Ouyang, Shuyin and Zhang, Jie M. and Sun, Zeyu and Penuela, Albert Merono },
booktitle = { 2025 IEEE/ACM 47th International Conference on Software Engineering (ICSE) },
title = {{ Knowledge-Enhanced Program Repair for Data Science Code }},
year = {2025},
volume = {},
ISSN = {},
pages = {898-910},
doi = {10.1109/ICSE55347.2025.00246},
url = {https://doi.ieeecomputersociety.org/10.1109/ICSE55347.2025.00246},
publisher = {IEEE Computer Society},
address = {Los Alamitos, CA, USA},
month =May}

@misc{ExpeRepair,
      title={EXPEREPAIR: Dual-Memory Enhanced LLM-based Repository-Level Program Repair}, 
      author={Fangwen Mu and Junjie Wang and Lin Shi and Song Wang and Shoubin Li and Qing Wang},
      year={2025},
      eprint={2506.10484},
      archivePrefix={arXiv},
      primaryClass={cs.SE},
      url={https://arxiv.org/abs/2506.10484}, 
}

@misc{SWE-bench,
      title={SWE-bench: Can Language Models Resolve Real-World GitHub Issues?}, 
      author={Carlos E. Jimenez and John Yang and Alexander Wettig and Shunyu Yao and Kexin Pei and Ofir Press and Karthik Narasimhan},
      year={2024},
      eprint={2310.06770},
      archivePrefix={arXiv},
      primaryClass={cs.CL},
      url={https://arxiv.org/abs/2310.06770}, 
}

@inproceedings{ref03_swebench_multimodal_2024,
  author = {Yang, John and Jimenez, Carlos E and Zhang, Alex and Lieret, Kilian and Yang, Joyce and Wu, Xindi and Press, Ori and Muennighoff, Niklas and Synnaeve, Gabriel and Narasimhan, Karthik and Yang, Diyi and Wang, Sida and Press, Ofir},
  booktitle = {International Conference on Learning Representations},
  editor = {Y. Yue and A. Garg and N. Peng and F. Sha and R. Yu},
  pages = {2794--2829},
  title = {SWE-bench Multimodal: Do AI Systems Generalize to Visual Software Domains?},
  url = {https://proceedings.iclr.cc/paper_files/paper/2025/file/07d6332ae36730707fddddba736d7b6c-Paper-Conference.pdf},
  volume = {2025},
  year = {2025}
}

@inproceedings{ref04_multi_swebench_2025,
  title={Multi-{SWE}-bench: A Multilingual Benchmark for Issue Resolving},
  author={Daoguang Zan and Zhirong Huang and Wei Liu and Hanwu Chen and Shulin Xin and Linhao Zhang and Qi Liu and Aoyan Li and Lu Chen and Xiaojian Zhong and Siyao Liu and Yongsheng Xiao and Liangqiang Chen and Yuyu Zhang and Jing Su and Tianyu Liu and RUI LONG and Ming Ding and liang xiang},
  booktitle={The Thirty-ninth Annual Conference on Neural Information Processing Systems Datasets and Benchmarks Track},
  year={2025},
  url={https://openreview.net/forum?id=MhBZzkz4h9}
}

@misc{ref05_swebench_pro_2025,
  title={SWE-Bench Pro: Can AI Agents Solve Long-Horizon Software Engineering Tasks?}, 
  author={Xiang Deng and Jeff Da and Edwin Pan and Yannis Yiming He and Charles Ide and Kanak Garg and Niklas Lauffer and Andrew Park and Nitin Pasari and Chetan Rane and Karmini Sampath and Maya Krishnan and Srivatsa Kundurthy and Sean Hendryx and Zifan Wang and Vijay Bharadwaj and Jeff Holm and Raja Aluri and Chen Bo Calvin Zhang and Noah Jacobson and Bing Liu and Brad Kenstler},
  year={2025},
  eprint={2509.16941},
  archivePrefix={arXiv},
  primaryClass={cs.SE},
  url={https://arxiv.org/abs/2509.16941}, 
}

@inproceedings{ref06_swebench_goes_live_2025,
  title={{SWE}-bench Goes Live!},
  author={Linghao Zhang and Shilin He and Chaoyun Zhang and Yu Kang and Bowen Li and Chengxing Xie and Junhao Wang and Maoquan Wang and Yufan Huang and Shengyu Fu and Elsie Nallipogu and Qingwei Lin and Yingnong Dang and Saravan Rajmohan and Dongmei Zhang},
  booktitle={The Thirty-ninth Annual Conference on Neural Information Processing Systems Datasets and Benchmarks Track},
  year={2025},
  url={https://openreview.net/forum?id=OGWkr7gXka}
}

@misc{ref07_swebench_memory_2025,
      title={Does SWE-Bench-Verified Test Agent Ability or Model Memory?}, 
      author={Thanosan Prathifkumar and Noble Saji Mathews and Meiyappan Nagappan},
      year={2025},
      eprint={2512.10218},
      archivePrefix={arXiv},
      primaryClass={cs.SE},
      url={https://arxiv.org/abs/2512.10218}, 
}

@misc{ref08_dissecting_leaderboards_2025,
      title={Dissecting the SWE-Bench Leaderboards: Profiling Submitters and Architectures of LLM- and Agent-Based Repair Systems}, 
      author={Matias Martinez and Xavier Franch},
      year={2026},
      eprint={2506.17208},
      archivePrefix={arXiv},
      primaryClass={cs.SE},
      url={https://arxiv.org/abs/2506.17208}, 
}

@misc{ref09_whats_in_benchmark_2026,
  title        = {What's in a Benchmark? The Case of SWE-Bench in Automated Program Repair},
  author       = {Matias Martinez and Xavier Franch},
  year         = {2026},
  eprint       = {2602.04449},
  archivePrefix= {arXiv},
  primaryClass = {cs.SE},
  url          = {https://arxiv.org/abs/2602.04449}
}

@misc{ref10_rigorous_eval_agents_2025,
      title={UTBoost: Rigorous Evaluation of Coding Agents on SWE-Bench}, 
      author={Boxi Yu and Yuxuan Zhu and Pinjia He and Daniel Kang},
      year={2025},
      eprint={2506.09289},
      archivePrefix={arXiv},
      primaryClass={cs.SE},
      url={https://arxiv.org/abs/2506.09289}, 
}

@misc{ref11_solved_issues_correctness_2025,
  title        = {Are "Solved Issues" in SWE-bench Really Solved?},
  author       = {You Wang and Michael Pradel and Zhongxin Liu},
  year         = {2025},
  eprint       = {2503.15223},
  archivePrefix= {arXiv},
  primaryClass = {cs.SE},
  url          = {https://arxiv.org/abs/2503.15223}
}

@misc{ref12_saving_swebench_2025,
      title={Saving SWE-Bench: A Benchmark Mutation Approach for Realistic Agent Evaluation}, 
      author={Spandan Garg and Benjamin Steenhoek and Yufan Huang},
      year={2026},
      eprint={2510.08996},
      archivePrefix={arXiv},
      primaryClass={cs.SE},
      url={https://arxiv.org/abs/2510.08996}, 
}

@misc{KGCompass,
      title={Enhancing Repository-Level Software Repair via Repository-Aware Knowledge Graphs}, 
      author={Boyang Yang and Haoye Tian and Jiadong Ren and Shunfu Jin and Yang Liu and Feng Liu and Bach Le},
      year={2025},
      eprint={2503.21710},
      archivePrefix={arXiv},
      primaryClass={cs.SE},
      url={https://arxiv.org/abs/2503.21710}, 
}

@inproceedings{SWE-Agent,
 author = {Yang, John and Jimenez, Carlos E. and Wettig, Alexander and Lieret, Kilian and Yao, Shunyu and Narasimhan, Karthik and Press, Ofir},
 booktitle = {Advances in Neural Information Processing Systems},
 editor = {A. Globerson and L. Mackey and D. Belgrave and A. Fan and U. Paquet and J. Tomczak and C. Zhang},
 pages = {50528--50652},
 publisher = {Curran Associates, Inc.},
 title = {SWE-agent: Agent-Computer Interfaces Enable Automated Software Engineering},
 url = {https://proceedings.neurips.cc/paper_files/paper/2024/file/5a7c947568c1b1328ccc5230172e1e7c-Paper-Conference.pdf},
 volume = {37},
 year = {2024}
}

@article{Agentless,
author = {Xia, Chunqiu Steven and Deng, Yinlin and Dunn, Soren and Zhang, Lingming},
title = {Demystifying LLM-Based Software Engineering Agents},
year = {2025},
issue_date = {July 2025},
publisher = {Association for Computing Machinery},
address = {New York, NY, USA},
volume = {2},
number = {FSE},
url = {https://doi.org/10.1145/3715754},
doi = {10.1145/3715754},
journal = {Proc. ACM Softw. Eng.},
month = jun,
articleno = {FSE037},
numpages = {24}
}

@misc{deepseekv3,
      title={DeepSeek-V3 Technical Report}, 
      author={DeepSeek-AI and Aixin Liu and Bei Feng and Bing Xue and Bingxuan Wang and Bochao Wu and Chengda Lu and Chenggang Zhao and Chengqi Deng and Chenyu Zhang and Chong Ruan and Damai Dai and Daya Guo and Dejian Yang and Deli Chen and Dongjie Ji and Erhang Li and Fangyun Lin and Fucong Dai and Fuli Luo and Guangbo Hao and Guanting Chen and Guowei Li and H. Zhang and Han Bao and Hanwei Xu and Haocheng Wang and Haowei Zhang and Honghui Ding and Huajian Xin and Huazuo Gao and Hui Li and Hui Qu and J. L. Cai and Jian Liang and Jianzhong Guo and Jiaqi Ni and Jiashi Li and Jiawei Wang and Jin Chen and Jingchang Chen and Jingyang Yuan and Junjie Qiu and Junlong Li and Junxiao Song and Kai Dong and Kai Hu and Kaige Gao and Kang Guan and Kexin Huang and Kuai Yu and Lean Wang and Lecong Zhang and Lei Xu and Leyi Xia and Liang Zhao and Litong Wang and Liyue Zhang and Meng Li and Miaojun Wang and Mingchuan Zhang and Minghua Zhang and Minghui Tang and Mingming Li and Ning Tian and Panpan Huang and Peiyi Wang and Peng Zhang and Qiancheng Wang and Qihao Zhu and Qinyu Chen and Qiushi Du and R. J. Chen and R. L. Jin and Ruiqi Ge and Ruisong Zhang and Ruizhe Pan and Runji Wang and Runxin Xu and Ruoyu Zhang and Ruyi Chen and S. S. Li and Shanghao Lu and Shangyan Zhou and Shanhuang Chen and Shaoqing Wu and Shengfeng Ye and Shengfeng Ye and Shirong Ma and Shiyu Wang and Shuang Zhou and Shuiping Yu and Shunfeng Zhou and Shuting Pan and T. Wang and Tao Yun and Tian Pei and Tianyu Sun and W. L. Xiao and Wangding Zeng and Wanjia Zhao and Wei An and Wen Liu and Wenfeng Liang and Wenjun Gao and Wenqin Yu and Wentao Zhang and X. Q. Li and Xiangyue Jin and Xianzu Wang and Xiao Bi and Xiaodong Liu and Xiaohan Wang and Xiaojin Shen and Xiaokang Chen and Xiaokang Zhang and Xiaosha Chen and Xiaotao Nie and Xiaowen Sun and Xiaoxiang Wang and Xin Cheng and Xin Liu and Xin Xie and Xingchao Liu and Xingkai Yu and Xinnan Song and Xinxia Shan and Xinyi Zhou and Xinyu Yang and Xinyuan Li and Xuecheng Su and Xuheng Lin and Y. K. Li and Y. Q. Wang and Y. X. Wei and Y. X. Zhu and Yang Zhang and Yanhong Xu and Yanhong Xu and Yanping Huang and Yao Li and Yao Zhao and Yaofeng Sun and Yaohui Li and Yaohui Wang and Yi Yu and Yi Zheng and Yichao Zhang and Yifan Shi and Yiliang Xiong and Ying He and Ying Tang and Yishi Piao and Yisong Wang and Yixuan Tan and Yiyang Ma and Yiyuan Liu and Yongqiang Guo and Yu Wu and Yuan Ou and Yuchen Zhu and Yuduan Wang and Yue Gong and Yuheng Zou and Yujia He and Yukun Zha and Yunfan Xiong and Yunxian Ma and Yuting Yan and Yuxiang Luo and Yuxiang You and Yuxuan Liu and Yuyang Zhou and Z. F. Wu and Z. Z. Ren and Zehui Ren and Zhangli Sha and Zhe Fu and Zhean Xu and Zhen Huang and Zhen Zhang and Zhenda Xie and Zhengyan Zhang and Zhewen Hao and Zhibin Gou and Zhicheng Ma and Zhigang Yan and Zhihong Shao and Zhipeng Xu and Zhiyu Wu and Zhongyu Zhang and Zhuoshu Li and Zihui Gu and Zijia Zhu and Zijun Liu and Zilin Li and Ziwei Xie and Ziyang Song and Ziyi Gao and Zizheng Pan},
      year={2025},
      eprint={2412.19437},
      archivePrefix={arXiv},
      primaryClass={cs.CL},
      url={https://arxiv.org/abs/2412.19437}, 
}

@inproceedings{lewis2020_rag,
author = {Lewis, Patrick and Perez, Ethan and Piktus, Aleksandra and Petroni, Fabio and Karpukhin, Vladimir and Goyal, Naman and K\"{u}ttler, Heinrich and Lewis, Mike and Yih, Wen-tau and Rockt\"{a}schel, Tim and Riedel, Sebastian and Kiela, Douwe},
title = {Retrieval-augmented generation for knowledge-intensive NLP tasks},
year = {2020},
isbn = {9781713829546},
publisher = {Curran Associates Inc.},
address = {Red Hook, NY, USA},
booktitle = {Proceedings of the 34th International Conference on Neural Information Processing Systems},
articleno = {793},
numpages = {16},
location = {Vancouver, BC, Canada},
series = {NIPS '20}
}

@article{ref37_faiss_2024,
  title={The faiss library},
  author={Douze, Matthijs and Guzhva, Alexandr and Deng, Chengqi and Johnson, Jeff and Szilvasy, Gergely and Mazar{\'e}, Pierre-Emmanuel and Lomeli, Maria and Hosseini, Lucas and J{\'e}gou, Herv{\'e}},
  journal={IEEE Transactions on Big Data},
  year={2025},
  publisher={IEEE}
}

@article{ref39_bm25_2009,
author = {Robertson, Stephen and Zaragoza, Hugo},
title = {The Probabilistic Relevance Framework: BM25 and Beyond},
year = {2009},
issue_date = {April 2009},
publisher = {Now Publishers Inc.},
address = {Hanover, MA, USA},
volume = {3},
number = {4},
issn = {1554-0669},
url = {https://doi.org/10.1561/1500000019},
doi = {10.1561/1500000019},
journal = {Found. Trends Inf. Retr.},
month = apr,
pages = {333–389},
numpages = {57}
}

@article{liu2023_lostinthemiddle,
    title = "Lost in the Middle: How Language Models Use Long Contexts",
    author = "Liu, Nelson F.  and
      Lin, Kevin  and
      Hewitt, John  and
      Paranjape, Ashwin  and
      Bevilacqua, Michele  and
      Petroni, Fabio  and
      Liang, Percy",
    journal = "Transactions of the Association for Computational Linguistics",
    volume = "12",
    year = "2024",
    address = "Cambridge, MA",
    publisher = "MIT Press",
    url = "https://aclanthology.org/2024.tacl-1.9/",
    doi = "10.1162/tacl_a_00638",
    pages = "157--173"
}

@inproceedings{zhang2023_repocoder,
    title = "{R}epo{C}oder: Repository-Level Code Completion Through Iterative Retrieval and Generation",
    author = "Zhang, Fengji  and
      Chen, Bei  and
      Zhang, Yue  and
      Keung, Jacky  and
      Liu, Jin  and
      Zan, Daoguang  and
      Mao, Yi  and
      Lou, Jian-Guang  and
      Chen, Weizhu",
    editor = "Bouamor, Houda  and
      Pino, Juan  and
      Bali, Kalika",
    booktitle = "Proceedings of the 2023 Conference on Empirical Methods in Natural Language Processing",
    month = dec,
    year = "2023",
    address = "Singapore",
    publisher = "Association for Computational Linguistics",
    url = "https://aclanthology.org/2023.emnlp-main.151/",
    doi = "10.18653/v1/2023.emnlp-main.151",
    pages = "2471--2484"
}

@inproceedings{zhang2024_autocoderover,
author = {Zhang, Yuntong and Ruan, Haifeng and Fan, Zhiyu and Roychoudhury, Abhik},
title = {AutoCodeRover: Autonomous Program Improvement},
year = {2024},
isbn = {9798400706127},
publisher = {Association for Computing Machinery},
address = {New York, NY, USA},
url = {https://doi.org/10.1145/3650212.3680384},
doi = {10.1145/3650212.3680384},
booktitle = {Proceedings of the 33rd ACM SIGSOFT International Symposium on Software Testing and Analysis},
pages = {1592–1604},
numpages = {13},
location = {Vienna, Austria},
series = {ISSTA 2024}
}

@article{chen2021_codex,
      title={Evaluating Large Language Models Trained on Code}, 
      author={Mark Chen and Jerry Tworek and Heewoo Jun and Qiming Yuan and Henrique Ponde de Oliveira Pinto and Jared Kaplan and Harri Edwards and Yuri Burda and Nicholas Joseph and Greg Brockman and Alex Ray and Raul Puri and Gretchen Krueger and Michael Petrov and Heidy Khlaaf and Girish Sastry and Pamela Mishkin and Brooke Chan and Scott Gray and Nick Ryder and Mikhail Pavlov and Alethea Power and Lukasz Kaiser and Mohammad Bavarian and Clemens Winter and Philippe Tillet and Felipe Petroski Such and Dave Cummings and Matthias Plappert and Fotios Chantzis and Elizabeth Barnes and Ariel Herbert-Voss and William Hebgen Guss and Alex Nichol and Alex Paino and Nikolas Tezak and Jie Tang and Igor Babuschkin and Suchir Balaji and Shantanu Jain and William Saunders and Christopher Hesse and Andrew N. Carr and Jan Leike and Josh Achiam and Vedant Misra and Evan Morikawa and Alec Radford and Matthew Knight and Miles Brundage and Mira Murati and Katie Mayer and Peter Welinder and Bob McGrew and Dario Amodei and Sam McCandlish and Ilya Sutskever and Wojciech Zaremba},
      year={2021},
      eprint={2107.03374},
      archivePrefix={arXiv},
      primaryClass={cs.LG},
      url={https://arxiv.org/abs/2107.03374}, 
}

@inproceedings{kim2013_par,
author = {Kim, Dongsun and Nam, Jaechang and Song, Jaewoo and Kim, Sunghun},
title = {Automatic patch generation learned from human-written patches},
year = {2013},
isbn = {9781467330763},
publisher = {IEEE Press},
booktitle = {Proceedings of the 2013 International Conference on Software Engineering},
pages = {802–811},
numpages = {10},
location = {San Francisco, CA, USA},
series = {ICSE '13}
}

@article{monperrus2018_automatic_software_repair_bibliography,
author = {Monperrus, Martin},
title = {Automatic Software Repair: A Bibliography},
year = {2018},
issue_date = {January 2019},
publisher = {Association for Computing Machinery},
address = {New York, NY, USA},
volume = {51},
number = {1},
issn = {0360-0300},
url = {https://doi.org/10.1145/3105906},
doi = {10.1145/3105906},
journal = {ACM Comput. Surv.},
month = jan,
articleno = {17},
numpages = {24}
}

@article{gazzola2019_automatic_software_repair,
author = {Gazzola, Luca and Micucci, Daniela and Mariani, Leonardo},
title = {Automatic Software Repair: A Survey},
year = {2019},
issue_date = {January 2019},
publisher = {IEEE Press},
volume = {45},
number = {1},
issn = {0098-5589},
url = {https://doi.org/10.1109/TSE.2017.2755013},
doi = {10.1109/TSE.2017.2755013},
journal = {IEEE Trans. Softw. Eng.},
month = jan,
pages = {34–67},
numpages = {34}
}

@INPROCEEDINGS{thien2013_semfix,
  author={Nguyen, Hoang Duong Thien and Qi, Dawei and Roychoudhury, Abhik and Chandra, Satish},
  booktitle={2013 35th International Conference on Software Engineering (ICSE)}, 
  title={SemFix: Program repair via semantic analysis}, 
  year={2013},
  volume={},
  number={},
  pages={772-781},
  doi={10.1109/ICSE.2013.6606623}}

@article{goues2012_genprog,
author = {Le Goues, Claire and Nguyen, ThanhVu and Forrest, Stephanie and Weimer, Westley},
title = {GenProg: A Generic Method for Automatic Software Repair},
year = {2012},
issue_date = {January 2012},
publisher = {IEEE Press},
volume = {38},
number = {1},
issn = {0098-5589},
url = {https://doi.org/10.1109/TSE.2011.104},
doi = {10.1109/TSE.2011.104},
journal = {IEEE Trans. Softw. Eng.},
month = jan,
pages = {54–72},
numpages = {19}
}

@misc{swebench_lite_website,
  title  = {SWE-bench Lite},
  author = {{SWE-bench Team}},
  year   = {2024},
  url    = {https://www.swebench.com/lite.html},
  note   = {Accessed 2026-02-18}
}

@misc{ContextBench2026,
      title={ContextBench: A Benchmark for Context Retrieval in Coding Agents}, 
      author={Han Li and Letian Zhu and Bohan Zhang and Rili Feng and Jiaming Wang and Yue Pan and Earl T. Barr and Federica Sarro and Zhaoyang Chu and He Ye},
      year={2026},
      eprint={2602.05892},
      archivePrefix={arXiv},
      primaryClass={cs.LG},
      url={https://arxiv.org/abs/2602.05892}, 
}

@misc{ref16_swe_contextbench_2026,
      title={SWE Context Bench: A Benchmark for Context Learning in Coding}, 
      author={Jared Zhu and Minhao Hu and Junde Wu},
      year={2026},
      eprint={2602.08316},
      archivePrefix={arXiv},
      primaryClass={cs.SE},
      url={https://arxiv.org/abs/2602.08316}, 
}

@article{ref47_mcnemar_1947,
  title={Note on the sampling error of the difference between correlated proportions or percentages},
  author={McNemar, Quinn},
  journal={Psychometrika},
  volume={12},
  number={2},
  pages={153--157},
  year={1947},
  publisher={Springer-Verlag}
}

@article{ref48_wilson_1927,
 ISSN = {01621459, 1537274X},
 URL = {http://www.jstor.org/stable/2276774},
 author = {Edwin B. Wilson},
 journal = {Journal of the American Statistical Association},
 number = {158},
 pages = {209--212},
 publisher = {[American Statistical Association, Taylor & Francis, Ltd.]},
 title = {Probable Inference, the Law of Succession, and Statistical Inference},
 urldate = {2026-03-16},
 volume = {22},
 year = {1927}
}

@article{ref49_bcd_2001,
 ISSN = {08834237, 21688745},
 URL = {http://www.jstor.org/stable/2676784},
 author = {Lawrence D. Brown and T. Tony Cai and Anirban DasGupta},
 journal = {Statistical Science},
 number = {2},
 pages = {101--117},
 publisher = {Institute of Mathematical Statistics},
 title = {Interval Estimation for a Binomial Proportion},
 urldate = {2026-03-16},
 volume = {16},
 year = {2001}
}

@misc{RepoFixEval2026,
    title={RepoFixEval: A Repository-Level Program Repair Benchmark From Issue Discovering to Bug Fixing},
    author={Tao Sun and Yang Yang and Xianfu Cheng and Jian Yang and Yintong Huo and Zhuoren Ye and Rubing Yang and Xiangyuan Guan and Wei Zhang and Hangyuan Ji and Changyu Ren and Mengdi Zhang and Xunliang Cai and Zhoujun Li},
    year={2024},
    url={https://openreview.net/forum?id=LaNCeNmoHR}
}

@misc{RGFL2026,
      title={RGFL: Reasoning Guided Fault Localization for Automated Program Repair Using Large Language Models}, 
      author={Melika Sepidband and Hamed Taherkhani and Hung Viet Pham and Hadi Hemmati},
      year={2026},
      eprint={2601.18044},
      archivePrefix={arXiv},
      primaryClass={cs.SE},
      url={https://arxiv.org/abs/2601.18044}, 
}

@misc{ref20_openhands_2024,
      title={OpenHands: An Open Platform for AI Software Developers as Generalist Agents}, 
      author={Xingyao Wang and Boxuan Li and Yufan Song and Frank F. Xu and Xiangru Tang and Mingchen Zhuge and Jiayi Pan and Yueqi Song and Bowen Li and Jaskirat Singh and Hoang H. Tran and Fuqiang Li and Ren Ma and Mingzhang Zheng and Bill Qian and Yanjun Shao and Niklas Muennighoff and Yizhe Zhang and Binyuan Hui and Junyang Lin and Robert Brennan and Hao Peng and Heng Ji and Graham Neubig},
      year={2025},
      eprint={2407.16741},
      archivePrefix={arXiv},
      primaryClass={cs.SE},
      url={https://arxiv.org/abs/2407.16741}, 
}

@misc{ref23_sgagent_2026,
      title={SGAgent: Suggestion-Guided LLM-Based Multi-Agent Framework for Repository-Level Software Repair}, 
      author={Quanjun Zhang and Chengyu Gao and Yu Han and Ye Shang and Chunrong Fang and Zhenyu Chen and Liang Xiao},
      year={2026},
      eprint={2602.23647},
      archivePrefix={arXiv},
      primaryClass={cs.SE},
      url={https://arxiv.org/abs/2602.23647}, 
}

@misc{ref25_reporepair_2026,
      title={RepoRepair: Leveraging Code Documentation for Repository-Level Automated Program Repair}, 
      author={Zhongqiang Pan and Chuanyi Li and Wenkang Zhong and Yi Feng and Bin Luo and Vincent Ng},
      year={2026},
      eprint={2603.01048},
      archivePrefix={arXiv},
      primaryClass={cs.SE},
      url={https://arxiv.org/abs/2603.01048}, 
}

@misc{jinaembeddings28192token,
      title={Jina Embeddings 2: 8192-Token General-Purpose Text Embeddings for Long Documents}, 
      author={Michael Günther and Jackmin Ong and Isabelle Mohr and Alaeddine Abdessalem and Tanguy Abel and Mohammad Kalim Akram and Susana Guzman and Georgios Mastrapas and Saba Sturua and Bo Wang and Maximilian Werk and Nan Wang and Han Xiao},
      year={2024},
      eprint={2310.19923},
      archivePrefix={arXiv},
      primaryClass={cs.CL},
      url={https://arxiv.org/abs/2310.19923}, 
}

@software{sourcecode,
  author       = {Anonymous Authors},
  title        = {Anonymous Artifact for ASE 2026 (Code and Data)},
  month        = mar,
  year         = 2026,
  publisher    = {Zenodo},
  doi          = {10.5281/zenodo.19112800},
  url          = {https://doi.org/10.5281/zenodo.19112800},
}

@misc{openai2025gpt41,
  title        = {Introducing GPT-4.1 in the API},
  author       = {{OpenAI}},
  year         = {2025},
  month        = apr,
  howpublished = {OpenAI product announcement},
  url          = {https://openai.com/index/gpt-4-1/},
  note         = {Accessed 2026-03-12}
}

@misc{swebench_leaderboards,
  title        = {SWE-bench Official Leaderboards},
  author       = {{SWE-bench Team}},
  year         = {2025},
  howpublished = {Online leaderboard},
  url          = {https://www.swebench.com/},
  note         = {Accessed 2026-02-18}
}

@misc{yang2025surveyllmbasedautomatedprogram,
      title={A Survey of LLM-based Automated Program Repair: Taxonomies, Design Paradigms, and Applications}, 
      author={Boyang Yang and Zijian Cai and Fengling Liu and Bach Le and Lingming Zhang and Tegawendé F. Bissyandé and Yang Liu and Haoye Tian},
      year={2025},
      eprint={2506.23749},
      archivePrefix={arXiv},
      primaryClass={cs.SE},
      url={https://arxiv.org/abs/2506.23749}, 
}

@misc{wu2023largelanguagemodelsfault,
      title={Large Language Models in Fault Localisation}, 
      author={Yonghao Wu and Zheng Li and Jie M. Zhang and Mike Papadakis and Mark Harman and Yong Liu},
      year={2023},
      eprint={2308.15276},
      archivePrefix={arXiv},
      primaryClass={cs.SE},
      url={https://arxiv.org/abs/2308.15276}, 
}

@article{10.1007/s10515-025-00549-x,
	author = {Wang, Bo and Deng, Ming and Chen, Mingda and Lin, Youfang and Zhou, Jianyi and Zhang, Jie M.},
	date = {2025/10/28},
	doi = {10.1007/s10515-025-00549-x},
	id = {Wang2025},
	isbn = {1573-7535},
	journal = {Automated Software Engineering},
	number = {1},
	pages = {26},
	title = {Assessing the effectiveness of recent closed-source large language models in fault localization and automated program repair},
	url = {https://doi.org/10.1007/s10515-025-00549-x},
	volume = {33},
	year = {2025}}

@article{Hanna2025RLMutationAPR,
      title={Reinforcement Learning for Mutation Operator Selection in Automated Program Repair}, 
      author={Carol Hanna and Aymeric Blot and Justyna Petke},
      year={2024},
      eprint={2306.05792},
      archivePrefix={arXiv},
      primaryClass={cs.SE},
      url={https://arxiv.org/abs/2306.05792}, 
}

@article{Kang2024LLMExplainableFaultLocalization,
   title={A Quantitative and Qualitative Evaluation of LLM-Based Explainable Fault Localization},
   volume={1},
   ISSN={2994-970X},
   url={http://dx.doi.org/10.1145/3660771},
   DOI={10.1145/3660771},
   number={FSE},
   journal={Proceedings of the ACM on Software Engineering},
   publisher={Association for Computing Machinery (ACM)},
   author={Kang, Sungmin and An, Gabin and Yoo, Shin},
   year={2024},
   month=jul, pages={1424–1446} 
}

@misc{Tao2025RetrievalAugmentedCodeGenerationSurvey,
      title={Retrieval-Augmented Code Generation: A Survey with Focus on Repository-Level Approaches}, 
      author={Yicheng Tao and Yao Qin and Yepang Liu},
      year={2026},
      eprint={2510.04905},
      archivePrefix={arXiv},
      primaryClass={cs.SE},
      url={https://arxiv.org/abs/2510.04905}, 
}

@misc{Martinez2014Astor,
      title={ASTOR: Evolutionary Automatic Software Repair for Java}, 
      author={Matias Martinez and Martin Monperrus},
      year={2014},
      eprint={1410.6651},
      archivePrefix={arXiv},
      primaryClass={cs.SE},
      url={https://arxiv.org/abs/1410.6651}, 
}

@article{Martinez2017RealBugs,
author = {Martinez, Matias and Durieux, Thomas and Sommerard, Romain and Xuan, Jifeng and Monperrus, Martin},
title = {Automatic repair of real bugs in java: a large-scale experiment on the defects4j dataset},
year = {2017},
issue_date = {August    2017},
publisher = {Kluwer Academic Publishers},
address = {USA},
volume = {22},
number = {4},
issn = {1382-3256},
url = {https://doi.org/10.1007/s10664-016-9470-4},
doi = {10.1007/s10664-016-9470-4},
journal = {Empirical Softw. Engg.},
month = aug,
pages = {1936–1964},
numpages = {29}
}

@INPROCEEDINGS{Liu2019FLBias,
  author={Liu, Kui and Koyuncu, Anil and Bissyandé, Tegawendé F. and Kim, Dongsun and Klein, Jacques and Le Traon, Yves},
  booktitle={2019 12th IEEE Conference on Software Testing, Validation and Verification (ICST)}, 
  title={You Cannot Fix What You Cannot Find! An Investigation of Fault Localization Bias in Benchmarking Automated Program Repair Systems}, 
  year={2019},
  volume={},
  number={},
  pages={102-113},
  doi={10.1109/ICST.2019.00020}}

@inproceedings{Mechtaev2016Angelix,
author = {Mechtaev, Sergey and Yi, Jooyong and Roychoudhury, Abhik},
title = {Angelix: scalable multiline program patch synthesis via symbolic analysis},
year = {2016},
isbn = {9781450339001},
publisher = {Association for Computing Machinery},
address = {New York, NY, USA},
url = {https://doi.org/10.1145/2884781.2884807},
doi = {10.1145/2884781.2884807},
booktitle = {Proceedings of the 38th International Conference on Software Engineering},
pages = {691–701},
numpages = {11},
location = {Austin, Texas},
series = {ICSE '16}
}

@inproceedings{Kang2022PatchPrioritization,
author = {Kang, Sungmin and Yoo, Shin},
title = {Language models can prioritize patches for practical program patching},
year = {2022},
isbn = {9781450392853},
publisher = {Association for Computing Machinery},
address = {New York, NY, USA},
url = {https://doi.org/10.1145/3524459.3527343},
doi = {10.1145/3524459.3527343},
booktitle = {Proceedings of the Third International Workshop on Automated Program Repair},
pages = {8–15},
numpages = {8},
location = {Pittsburgh, Pennsylvania},
series = {APR '22}
}

@article{Zeller2002DeltaDebugging,
author = {Zeller, Andreas and Hildebrandt, Ralf},
title = {Simplifying and Isolating Failure-Inducing Input},
year = {2002},
issue_date = {February 2002},
publisher = {IEEE Press},
volume = {28},
number = {2},
issn = {0098-5589},
url = {https://doi.org/10.1109/32.988498},
doi = {10.1109/32.988498},
journal = {IEEE Trans. Softw. Eng.},
month = feb,
pages = {183–200},
numpages = {18}
}

@article{Abreu2009SBFL,
author = {Abreu, Rui and Zoeteweij, Peter and Golsteijn, Rob and van Gemund, Arjan J. C.},
title = {A practical evaluation of spectrum-based fault localization},
year = {2009},
issue_date = {November, 2009},
publisher = {Elsevier Science Inc.},
address = {USA},
volume = {82},
number = {11},
issn = {0164-1212},
url = {https://doi.org/10.1016/j.jss.2009.06.035},
doi = {10.1016/j.jss.2009.06.035},
journal = {J. Syst. Softw.},
month = nov,
pages = {1780–1792},
numpages = {13}
}

@article{Wotawa2002MBDvsSlicing,
author = {Wotawa, Franz},
title = {On the relationship between model-based debugging and program slicing},
year = {2002},
issue_date = {02/01/2002},
publisher = {Elsevier Science Publishers Ltd.},
address = {GBR},
volume = {135},
number = {1–2},
issn = {0004-3702},
url = {https://doi.org/10.1016/S0004-3702(01)00161-8},
doi = {10.1016/S0004-3702(01)00161-8},
journal = {Artif. Intell.},
month = feb,
pages = {125–143},
numpages = {19}
}
